\documentclass[preprint,review]{elsarticle} 
\usepackage[english]{babel}
\usepackage[latin1]{inputenc}

\usepackage{algorithm}
\usepackage{algpseudocode}

\usepackage{graphicx}
 
\usepackage[cmex10]{amsmath}

\usepackage{psfrag}

\usepackage{textcomp}

\usepackage{subfig}
\usepackage{wrapfig}
\usepackage{siunitx} 
\usepackage{float}
\usepackage{multirow}

\usepackage[usenames, dvipsnames]{color} 

\usepackage[english]{varioref}

\usepackage{booktabs}

\usepackage{gensymb}
\usepackage{amssymb}
\usepackage{fancyhdr}
\usepackage{braket} 

\usepackage{cleveref}

\usepackage{mathtools}

\usepackage[showframe=false]{geometry}

\usepackage{xcolor,colortbl}

\newcommand{\fnsnum}[1]{\footnotesize{\num{#1}}}

\newcommand{\TwoRowCell}[1]{ {\scriptsize \begin{tabular}{@{}c@{}} #1 \end{tabular}}  }

\newcommand{\parRP}{	\psfrag{a}{\tiny $\hat\mu_1$}
					\psfrag{b}{\tiny $\hat\mu_2$}
					\psfrag{c}{\tiny $\hat\mu_3$}
					\psfrag{w1}{\tiny $\hat\mu_4$}
					\psfrag{w2}{\tiny $\hat\mu_5$}
					\psfrag{tilt}{\tiny $\psi$}
					\psfrag{azimut}{\tiny $\zeta$}
				   }
\newcommand{\power}{\psfrag{Power [kW]}{\hspace*{2mm}\tiny{\si{\kilo\watt}}}
					\psfrag{time [day]}{\tiny{time (day)}}
				   }
\newcommand{\rmse}{	\psfrag{kW}[b]{\tiny{$RMSE_d$ (\si{\kilo\watt})}}
					\psfrag{day}{\tiny{time (day)}}
				  }

\definecolor{Red}{rgb}{1.0,0.6,0.6}
\definecolor{Blue}{rgb}{0.6,0.8,1.0}
\definecolor{Green}{rgb}{0.7,1.0,0.4}
\definecolor{Gray}{gray}{.85}
\newcolumntype{R}{>{\columncolor{Red}}c}
\newcolumntype{b}{>{\columncolor{Blue}}c}
\newcolumntype{g}{>{\columncolor{Green}}c}
\newcolumntype{G}{>{\columncolor{Gray}}c}

\DeclarePairedDelimiter{\abs}{\lvert}{\rvert} 

\journal{Solar Energy}

\begin{document}
\begin{frontmatter}
\title{Model Estimation for Solar Generation Forecasting \\ using Cloud Cover Data}
\author{Daniele Pepe}\ead{pepe@diism.unisi.it}              
\author{Gianni Bianchini}\ead{giannibi@diism.unisi.it}    
\author{Antonio Vicino}\ead{vicino@diism.unisi.it}  
\address {Dipartimento di Ingegneria dell'Informazione e Scienze Matematiche, Universit\`a di Siena, \\Via Roma 56, 53100 Siena, Italy}  

\begin{abstract}
This paper presents a parametric model approach to address the problem of photovoltaic generation forecasting in a scenario where measurements of meteorological variables, i.e., solar irradiance and temperature, are not available at the plant site. This scenario is relevant to electricity network operation, when a large number of PV plants are deployed in the grid. The proposed method makes use of raw cloud cover data provided by a meteorological service combined with power generation measurements, and is particularly suitable in PV plant integration on a large-scale basis, due to low model complexity and computational efficiency. An extensive validation is performed using both simulated and real data. 
\end{abstract}

\end{frontmatter}

\section{Introduction}
A major challenge in the integration of renewable energy sources into the grid is that power generation is intermittent, difficult to control, and strongly depending on the variation of weather conditions.
For these reasons, forecasting of renewable distributed generation has become a fundamental need to grid operators. In this respect, solar generation forecasts on multiple time horizons are needed to satisfy grid constraints and demand. In particular, short-term forecasts are required for the purposes of power plant operation,
grid balancing, real-time unit dispatching, automatic generation control, and energy trading. On the contrary, longer-term forecasts are of
interest to Distribution System Operators (DSO) and Transmission System Operators (TSO) for unit commitment, scheduling and
for improving balance area control (see, e.g., \cite{BIB:DENHOLM2007_1, BIB:DENHOLM2007_2, ALBUYEH09,CIRED12} and references therein).

Concerning solar power generation, much attention has been paid to the problem of obtaining accurate day-ahead and hour-ahead forecasts of solar irradiance and/or generated power (see \cite{diagne2013review,coimbraoverview} for a comprehensive overview on the subject). Most contributions focus on solar irradiance forecasting \cite{Inman2013535}.  Widely adopted approaches are based on Artificial Neural Networks (ANNs) \cite{CAPIZZI12,WU11,SHANXU10,BIB:CORNARO} or Support Vector Machines \cite{RAGNACCI12} with different types of input data.
Alternatively, classical linear time series forecasting methods are used in \cite{REIKARD09,LORENZ09,BACHER09}, where the considered time series is typically the global horizontal irradiance (GHI) normalized with a clear-sky model (see \cite{WONG01,ASHRAE09} for a comprehensive review). Global radiation forecasts are then fed along with temperature forecasts to a simulation model of the plant \cite{PATEL06} to compute the prediction of power generation. In any case, computing reliable generation forecasts from predicted meteorological variables hinges upon the availability of a reliable model of the plant, be it physical or estimated from data.

Unfortunately, in many practical scenarios, neither reliable plant models, nor direct on-site measurements of solar irradiance and other meteorological variables such as temperature are available. This is the typical case of a a DSO dealing with hundreds or thousands of distributed heterogeneous and independently-operated solar plants; in this case, the only available plant data is represented by generated power measurements provided by electronic meters. The contribution of this paper focuses on the problem of estimating reliable generation models in such limited information contexts.
In~\cite{BIB:ISGT,BIB:CDC13}, two methods are proposed to estimate the parameters of the PVUSA model~\cite{BIB:PVUSA} of a PV plant via a recursive framework based only on measures of generated power and temperature forecasts. This solution is not always data-efficient, since generation data collected during cloudy days are not exploited.

In order to obtain more accurate results, power measurements can be combined with further data coming from a weather service. Such data are typically averaged over large geographic zones and therefore they may be scarcely informative for a specific spot, yet they may provide useful information when the goal is to address the aggregation of multiple plants over a macro area. In this respect, cloud cover measurements derived from satellite imaging are an exception, since they can be made available for specific locations with good spatial and temporal resolution
(up to $2.5~\si{\kilo\meter}\times 2.5~\si{\kilo\meter}$ and 
$30~\si{\minute}$, respectively) \cite{HAMMER1999}.
Based on such data, suitable models can be estimated for irradiance forecasting purposes (see \cite{BIB:PEREZ2007,PEREZ2010,BIB:KANG2015} for details). Irradiance forecasts can be used as inputs to a plant generation model in order to provide energy production forecasts. Of course, the latter step requires that a reliable model of power generation from irradiance be previously estimated.

\enlargethispage{\baselineskip}
In this paper, we present a novel approach for direct forecasting of PV plant power generation from cloudiness data. To this purpose, a class of parametric models is introduced which efficiently exploits the PVUSA model~\cite{BIB:PVUSA} and the notion of Cloud Cover Factor (CCF)~\cite{BIB:KASTEN,BIB:KIMURA} in a limited information scenario. More specifically, the model parameters are estimated using generated power data combined with additional information provided by a meteorological service for the area where the plant is located. Such data consist of a time series of raw cloud cover and temperature reports. For the proposed models, estimation procedures based on Recursive Least Squares (RLS) and the Extended Kalman Filter (EKF)~\cite{BIB:KALMAN}, are devised. The properties and the performance of the proposed method are demonstrated both in a simulated scenario and on experimental data from a plant currently in operation. Preliminary results leading to this paper were presented in \cite{BIB:ENERGYCON16}. 

The manuscript is organized as follows. In Section~\ref{PRE} we introduce the modelling tools used; in Section~\ref{SEC:JOIN_MODEL} the proposed models are presented. Estimation procedures are presented in Section~\ref{SEC:IDENTIFICATION}. Section~\ref{sec:forecasting} addresses the relevant forecasting problems, while performance evaluation is discussed in Section~\ref{SEC:PEVA}. Simulation results are illustrated in Section~\ref{SEC:SIM_RES}. Experimental validation results and related discussion are reported in Section~\ref{SEC:EXP_RES}. Finally, conclusions are drawn in Section~\ref{SEC:CONC}.

\section{Models and methods}\label{PRE}
\subsection{The PVUSA photovoltaic plant model}
\label{SEC:PVUSA}
A PV plant can be efficiently modelled using the PVUSA model~\cite{BIB:PVUSA}, which expresses the instantaneous generated power as a function of irradiance and air temperature according to the equation:
\begin{equation}
	\label{EQ:PVUSA}
	P = \mu_1 I + \mu_2 I^2 + \mu_3 IT,
\end{equation}
where $P$, $I$, and $T$ are the generated power (\si{\kilo\watt}), irradiance (\si{\watt/\metre^{2}}), and air temperature (\si{\degreeCelsius}), respectively, and $\mu_1$, $\mu_2$, $\mu_3$, are the model parameters. \\
Model \eqref{EQ:PVUSA} is linear and parsimonious in terms of number of parameters. Despite its simplicity, very good accuracy is obtained when this model is fit to real measured data. Moreover, due to their reliability, temperature forecasts taken from a meteorological service can be used efficiently in place of actual measurements in order to estimate the model parameters, as shown in \cite{BIB:ISGT}. This cannot be done for irradiance, as forecasts are much less reliable.\\
We find it convenient to express \eqref{EQ:PVUSA} also in the form
\begin{equation}
	\label{EQ:PVUSAK}
	P = \mu_1 (1+ \eta_2 I + \eta_3 T)I,
\end{equation}
where $\mu_1$ plays the role of the main power/irradiance gain of the plant, while $\eta_2=\mu_2/\mu_1$ and $\eta_3= \mu_3/\mu_1$ introduce correction terms. It is worth noticing that $\eta_2$ and $\eta_3$ in \eqref{EQ:PVUSAK} are characterized by a narrow range of variability among different PV technologies. Indeed, typical values of these parameters are given by \cite{BIB:PVUSA}:
\begin{equation}
\begin{array}{l}
  \eta_2\in {\cal S}_2 = \left[\num{-2.5e-4}, \num{-1.9e-5}\right], \\
  \eta_3 \in {\cal S}_3 = \left[\num{-4.8e-3}, \num{-1.7e-3}\right].
\end{array} \label{EQ:AtoBC}
\end{equation}

\subsection{Cloud Cover Index and Cloud Cover Factor}
\label{SEC:CLOUD_COVERING}
The Cloud Cover Index (CCI), here denoted by $N$, refers to the fraction of the sky obscured by clouds when observed from a particular location. In meteorology, $N$ is measured in {\sl okta}~\cite{BIB:OKTA}, i.e., an integer ranging from \SI{0}{okta} (clear sky) up to \SI{8}{okta}  (completely overcast). Weather services often provide this information in terms of percentage of covered sky: from $0.0$ (clear sky) to $1.0$ (completely overcast) in steps of $0.1$. 

\begin{figure}
	\psfrag{N}{\tiny{$N$}}
	\psfrag{CCF1}{\tiny{$C(N)$}}
	\psfrag{CCF2}{\tiny{$C(N)$}}
	\centering
	\includegraphics[width=0.5\textwidth,height=4cm]{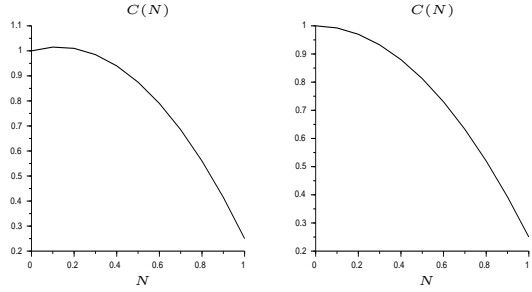}
	\caption{Qualitative behavior of the CCF $C(N)$ for $\mu_4>0$ (left) and $\mu_4<0$ (right).}
	\label{FIG:CCF_EXAMPLE}
\end{figure}

A widely investigated issue is the relationship between the CCI and the global irradiance on a surface. In~\cite{BIB:KIMURA}, a simple way of estimating solar radiation in cloudy days is to multiply the clear-sky irradiance by the so-called Cloud Cover Factor (CCF), which is defined as a function of the CCI. Let $I^0$ be the clear-sky irradiance on a given surface. It turns out that the global irradiance on the same surface is given by 
\begin{equation}
	\label{EQ:CCF}
	I(N) = C(N) \cdot I^0,
\end{equation}
where $C(N)$ is the CCF, which can be expressed in the form
\begin{equation}
	\label{EQ:KIMURA}
	C(N) = 1 + \mu_4 N + \mu_5 N^2 .
\end{equation}
In (\ref{EQ:KIMURA}), the parameters $\mu_4$ and $\mu_5$ depend on the local climate. In particular, $\mu_5$ is always negative. Depending on the sign of $\mu_4$, the curve $C(N)$ may be non-monotonic (see Fig.~\ref{FIG:CCF_EXAMPLE}).
Indeed, in some locations $C(N)$ may rise above 1 (i.e., the clear-sky value) under slightly cloudy sky (e.g, $0<N<0.3$), since the diffuse radiation becomes significant with middle-low cloudiness. Moreover, while the CCF is independent of solar elevation, it is subject to seasonal variations. Alternative CCF models exist in the literature~\cite{BIB:KASTEN}.

\enlargethispage*{\baselineskip}
The CCF is usually estimated for a given location using on-site irradiance measurements~\cite{BIB:KIM}. It is worth to remark, however, that the method presented in this paper does not rely on such measurements.
\subsection{Clear-sky irradiance model}

In this paper we need to exploit an estimate of the clear-sky irradiance on a given surface, as opposed to the GHI. As a first step, we make use of the Heliodon simulator model \cite{BIB:HELIODON}. This model allows to compute the theoretical clear-sky normal irradiance (\si{\watt/\meter^2}) from the solar altitude~$h$, i.e., the angle over the horizon (rads), as:
        \begin{equation}
                \label{EQ:ELIODON}
                I^{cs,n} = \left\{\begin{array}{ll} A\cdot0.7^{ \left( \frac{1}{\sin{h}} \right)^{0.678}}  & \mbox{if}~~0<h<\pi/2 \\ 0 & \mbox{otherwise,} \end{array}\right.
        \end{equation}
        where $A=1353$ \si{\watt/\meter^2} denotes the apparent extraterrestrial irradiance. Given the theoretical clear-sky normal irradiance $I^{cs,n}$, the clear-sky irradiance on an inclined surface $I^{cs}$ can be derived from $I^{cs,n}$ and surface orientation with respect to the sun position. Denoting by $\zeta$ the surface azimuth and $\psi$ the surface tilt angle, one has that
\begin{equation}
\label{EQ:ICS}
I^{cs}=\left[\sin(\psi)\cos(h)\cos(\zeta-\gamma)+\cos(\psi)\sin(h)\right] I^{cs,n},
\end{equation}
where $\gamma$ is the solar azimuth. Clearly, $I^{cs}$ can be computed for given values of $\zeta$ and $\psi$ from latitude, longitude and time of day. For $\psi=0$, the irradiance on an horizontal surface is obtained.

In this study, we make the realistic assumption that the exact orientation of the PV panel surfaces of the considered plant is not known a-priori. However, it is reasonable to assume that the plant is efficiently oriented for the specific latitude according to, e.g., the guidelines given in~\cite{BIB:ORIENTATION}. Therefore, we will use as the reference value of the theoretical clear-sky irradiance $I^0$ for a specific plant, the value of (\ref{EQ:ICS}) for $(\zeta,\psi)=(\zeta_0,\psi_0)$, where $\zeta_0$ and $\psi_0$ are taken from the above guidelines, i.e.,
\begin{equation}
\label{EQ:ICS0}
I^{0}=\left[\sin(\psi_0)\cos(h)\cos(\zeta_0-\gamma)+\cos(\psi_0)\sin(h)\right] I^{cs,n} .
\end{equation}
The choice of the model in \eqref{EQ:ELIODON} is mainly due to its simplicity. More accurate models taking into account atmospheric parameters such as aerosol optical depth,
ozone, and water vapor, could be used upon availability of such data in order to improve the performance of the forecasting method proposed in this work.

\section{Plant generation model}\label{SEC:JOIN_MODEL}
The plant model \eqref{EQ:PVUSA} and the irradiance model \eqref{EQ:CCF}$-$\eqref{EQ:KIMURA} introduced in the previous section can be combined in order to obtain an expression of the generated power as a function of $N$, $T$, and $I^{0}$, that is
\begin{equation}\label{EQ:COMBI}
	P = P(I^{0},T,N) = (\mu_1 + \mu_2I(N) +  \mu_3 T)I(N),
\end{equation}
where
\begin{equation}\label{EQ:COMBI2}	
	I(N) = \left( 1 + \mu_4 N + \mu_5 N^2 \right)I^{0}.
\end{equation}
Therefore, the overall model is defined by the parameter vector
$$
\mu = \begin{bmatrix} \mu_1 &  \mu_2 & \mu_3 & \mu_4 & \mu_5 \end{bmatrix}^T,
$$
and \eqref{EQ:COMBI}$-$\eqref{EQ:COMBI2} can be rewritten in a standard regression form as
\begin{equation}
	\label{EQ:COMP_P}
	P = \varphi^T(I^{0}, T, N)\theta(\mu),
\end{equation}
where $\theta(\mu)$ and $\varphi(I^{0}, T, N)$ are given by
\begin{equation}\label{EQ:LAMBDA}
\setcounter{MaxMatrixCols}{20}
\begin{array}{rcl}
\theta(\mu) & = & { \left[\begin{array}{ccccc}
								\theta_1(\mu) 					& 
								\theta_2(\mu) 						& 
								\theta_3(\mu) 						&
								\theta_4(\mu) 					&
								\theta_5(\mu) 				
								
					   \end{array}\right.} \\
					   & & {\left.\begin{array}{cccccc}
					   			\theta_6(\mu)  &
								\theta_7(\mu) 				&
								\theta_8(\mu)					&
								\theta_9(\mu) 						&
								\theta_{10}(\mu) 						&
								\theta_{11}(\mu)   
								\end{array}\right]^T} \\
~ & = & { \left[\begin{array}{ccccc}
								\mu_1 					& 
								\mu_1\mu_4 						& 
								\mu_1\mu_5 						&
								\mu_2 					&
								2\mu_2\mu_4 				
								
					   \end{array}\right.} \\
					   & & {\left.\begin{array}{cccccc}
					   			\mu_2\mu_4^2+2\mu_2\mu_5  &
								2\mu_2\mu_4\mu_5 				&
								\mu_2\mu_5^2					&
								\mu_3 						&
								\mu_3\mu_4 						&
								\mu_3\mu_5   
								\end{array}\right]^T,}

\end{array}
\end{equation}
\begin{equation}\label{EQ:PHI}
\setcounter{MaxMatrixCols}{20}
\begin{array}{l}
\varphi(I^0, T, N)  =  {\left[\begin{array}{ccccc}
							I^0 			& 
					I^0N 			& 
					I^0N^2 		& 
					{I^0}^2 		& 
					{I^0}^2N 		
					   \end{array}\right.} \\
					    {\left.\begin{array}{cccccc}
					   			{I^0}^2 N^2 	&
					{I^0}^2N^3 		&
					{I^0}^2N^4 		& 
					TI^0 			& 
					TI^0N 		& 
					TI^0N^2 
					\end{array}\right]^T} .
\end{array} 
\end{equation}
Notice that the model in~\eqref{EQ:COMP_P}-\eqref{EQ:PHI} is not linear in
the parameter vector $\mu$. 
In order to estimate model parameters, two approaches will be pursued in the following by comparing their respective performances:
\begin{itemize}
\item[(N)] Keep the parameter vector size to a minimum and derive a nonlinear estimation algorithm based on the Extended Kalman Filter (EKF);
\item[(L)] Introduce an overparameterization by disregarding the parameterization of $\theta(\mu)$ in $\mu$ and assuming $\theta(\cdot)$ to be a vector of independent parameters, thus making the model linear and amenable to standard least squares estimation.
\end{itemize}
The former approach has the advantage of keeping the model parsimonious and improving its identifiability features, while the latter is introduced with the purpose of making the estimation task simpler.

\subsection{Nonlinear (N) models}
\label{SSEC:EKF_APPROACH}
The first idea proposed here is to tackle the nonlinear estimation problem exploiting the EKF. To this purpose, let us consider model~\eqref{EQ:COMP_P}$-$\eqref{EQ:PHI}
for the following two choices of the independent parameter vector $\mu$:
\begin{enumerate}
\item N5 model
\begin{equation}
	\mu = \begin{bmatrix} \mu_1 & \mu_2 & \mu_3 & \mu_4 & \mu_5 \end{bmatrix}^T ,
\end{equation}
\begin{equation}\label{EQ:LAMBDAN5}
\setcounter{MaxMatrixCols}{20}
\begin{array}{rcl}
\theta(\mu) & = & { \left[\begin{array}{ccccc}
								\mu_1 					& 
								\mu_1\mu_4 						& 
								\mu_1\mu_5 						&
								\mu_2 					&
								2\mu_2\mu_4 				
								
					   \end{array}\right.} \\
					   & & {\left.\begin{array}{cccccc}
					   			\mu_2\mu_4^2+2\mu_2\mu_5  &
								2\mu_2\mu_4\mu_5 				&
								\mu_2\mu_5^2					&
								\mu_3 						&
								\mu_3\mu_4 						&
								\mu_3\mu_5   
								\end{array}\right]^T,}

\end{array}
\end{equation}

which is the original nonlinear model \eqref{EQ:COMP_P}$-$\eqref{EQ:PHI} characterized by a minimal number of parameters, and 
\item N6 model
\begin{equation}\label{EQ:MU6}
	\mu = \begin{bmatrix} \mu_1 & \mu_2 & \mu_3 & \mu_4 & \mu_5 & \mu_6 \end{bmatrix}^T,
\end{equation}
\begin{equation}\label{EQ:LAMBDAN6}
\setcounter{MaxMatrixCols}{20}
\begin{array}{rcl}
\theta(\mu) & = & { \left[\begin{array}{ccccc}
								\mu_1 					& 
								\mu_1\mu_4 						& 
								\mu_1\mu_5 						&
								\mu_2 					&
								2\mu_6 				
								
					   \end{array}\right.} \\
					   & & {\left.\begin{array}{cccccc}
					   			\mu_4\mu_6+2\mu_2\mu_5  &
								2\mu_5\mu_6 				&
								\mu_2\mu_5^2					&
								\mu_3 						&
								\mu_3\mu_4 						&
								\mu_3\mu_5   
								\end{array}\right]^T,}

\end{array}
\end{equation}
which implies a slight overparameterization with respect to model N5, amounting to treating $\mu_2$, $\mu_4$ and $\mu_2\mu_4=\mu_6$ as independent parameters. 
\end{enumerate}

For either choice, the model takes the form \eqref{EQ:COMP_P}
\begin{equation*}
	\label{EQ:COMP_P2}
	P = \varphi^T(I^{0}, T, N)\theta(\mu) .
\end{equation*}

\subsection{Linear overparameterized (L) model}
\label{SSEC:OP_APPROACH}
This model is derived by treating $\theta(\mu)$ in~\eqref{EQ:LAMBDA} as a vector $\theta$ of 11 independent parameters.
This way model \eqref{EQ:COMP_P} becomes
\begin{equation*}
	\label{EQ:OP_model}
	P = \varphi^T(I^0, T, N)\ \theta
\end{equation*}
and is clearly linear in the parameters. For this model, a standard Recursive Least Squares (RLS) algorithm can be used to estimate $\theta$.

\section{Model estimation}\label{SEC:IDENTIFICATION}
For the models introduced in Sections~\ref{SSEC:EKF_APPROACH} and \ref{SSEC:OP_APPROACH}, an on-line update of the estimate of the parameter vector $\mu$ (or $\theta$) is performed. 
To the purpose of illustrating the procedure, let us define the following quantities:
\begin{itemize}
\item $\tau_s$: sampling time (mins);
\item $k$: discrete time index;
\item $d$: generic day;
\item $\tau_d(k)$: time of day (TOD) corresponding to index $k$ (at the given longitude);
\item $I^0(k)$: theoretical clear-sky solar irradiance at time index $k$ at the plant site, computed according to (\ref{EQ:ELIODON})-(\ref{EQ:ICS0}); 
\item ${\cal K}_d=\{\underline k_d, \dots, \overline k_d\}$: set of time indices pertaining to light hours in day $d$, i.e., the theoretical irradiance \eqref{EQ:ELIODON}-\eqref{EQ:ICS0} is nonzero for $\tau_d(k)$ corresponding to $\underline k_d \leq k \leq \overline k_d$, and moreover $\underline k_{d+1}=\overline k_{d}+1$, that is, only daylight time indices are considered;
\item $\hat\mu(k)$ [$\hat\theta(k)$]: estimate of the parameter vector at time $k$, after processing the last data sample, being $\hat\mu(0)$ [$\hat\theta(0)$] the initial guess;
\item ${W}(k)$: weather report for time $k$ relative to the the plant area provided by a meteorological service. More specifically, $W(k)=\{N(k),T(k)\}$, where $N(k)$ and $T(k)$ are the cloud cover and temperature reports, respectively\footnote{$T(k)$ can be represented by either a real-time estimate of the plant site temperature or even some forecast thereof, computed in advance. Indeed, it has been shown \cite{BIB:ISGT} that the reliability of day-ahead temperature forecasts is comparable to that of actual measurements for model estimation purposes. Obviously, when on-site measurements of temperature $T^m(k)$ are available, then $T(k)$ should be taken equal to $T^m(k)$. In the experimental section of this paper, data from day-ahead forecasts are considered.}.
\item $P^m(k)$: measured generated power at time $k$;
\item $D(k)=\{P^m(k),W(k)\}=\{P^m(k),N(k),T(k)\}$: data available at time $k$;
\item $\phi(k) = \varphi(I^0(k),T(k),N(k)$): regression vector at time $k$ (see (\ref{EQ:PHI}));
%
%
%
\end{itemize}
The estimation procedure is recursive. At each time step $k$, a new power-weather data sample $\{P^m(k), W(k)\}$ is acquired, the algorithm is run and a current estimate $\hat{\mu}(k)$ [$\hat{\theta}(k)$] of the parameter vector is computed according to the criteria outlined below.
\begin{itemize}
\item {\sl N5 and N6 models}. For the purpose of estimating $\mu$ we consider the standard extended Kalman filter (EKF) formulation defined by the following state-space model \cite{BIB:KALMAN}:
\begin{subequations} \label{EQ:EKF}
	\begin{align}
		\mu(k+1) 	&= \mu(k) \label{EQ:KF1} \\
		P(k)			&= \phi^T(k)\theta(\mu(k)) + w(k), \label{EQ:KF2}
	\end{align}
\end{subequations}
where $\mu(k)$ represents the state vector, $P(k)$ is the output, and $w(k)$ is a zero-mean Gaussian white noise with given variance~$r$ which describes both the measurement noise and the model uncertainty. According to the standard EKF theory, the following update law for the parameter estimate $\hat\mu(k)$ is obtained:
\begin{equation}
\hat{\mu}(k)	= \hat{\mu}(k-1) + K(k) \left[ P^m(k) - \phi^T(k)\theta(\hat\mu(k-1)) \right],
\end{equation}
where
\begin{equation}
	\vspace*{-2mm} K(k) 			= \frac{R(k-1)H^T(k)}{H(k)R(k-1)H^T(k) + r},
\end{equation}
and the estimate covariance matrix $R(k)$ is updated at each step according to the rule	
\begin{equation}
	R(k) 			=  \left( I - H^T(k)K^T(k) \right) R(k-1), 
\end{equation}
being 
$$
H(k) = \frac{\partial}{\partial \mu} \left[ \phi^T(k) \theta(\mu) \right]_{\mu=\hat{\mu}(k-1)} .
$$
As in common practice, the initial estimate covariance matrix $R(0)$ is computed as $R(0) = l(0) \cdot \cal I$. The higher the confidence on the initial guess $\hat{\mu}(0)$, the lower $l(0)$ can be chosen.  The initialization of the parameter vector estimate $\hat\mu(0)$ is discussed at the end of this section.

\item {\sl L model}. A recursive least-squares (RLS) estimation step is performed, i.e., we compute
\begin{equation}
L(k) = V(k) \phi(k),
\end{equation}
where $V(k)$ is the weight matrix at time $k$, which in turn is recursively computed according to the standard RLS update law:
\begin{equation}
V(k) = V(k-1) - \frac{V(k-1) \phi(k) \phi^T(k) V(k-1)}{1 + \phi^T(k) V(k-1) \phi(k)}.
\end{equation}
Finally the current parameter estimate is updated according to the law
\begin{equation}
\hat{\theta}(k) = \hat{\theta}(k-1) + L(k) \left[P^m(k) - \phi^T(k)\hat{\theta}(k-1) \right].
\end{equation}
As in the usual RLS practice, the initial weight matrix $V(0)$ is set to $V(0)=l(0)\cdot~\cal I$, where $\cal I$ is the identity matrix and $l(0)>0$ is chosen according to the confidence given to the initial parameter guess $\hat\theta(0)$ (higher $l(0)$ meaning less confidence). 
\end{itemize}

As far as the initialization of the parameter vector is concerned, a reliable guess for the initial values $\hat\mu(0)$ in model N5 can be computed from the following guidelines.
\begin{itemize}
\item A good guess for the plant gain $\mu_1$ is given by $\hat\mu_1(0)=P_{nom}/1000$ where $P_{nom}$ is the nominal plant power in \si{\kilo\watt} \cite{BIB:ISGT};
\item the initial values $\hat\mu_2(0)$ and $\hat\mu_3(0)$ can be chosen such that $\eta_2(0)=\hat\mu_2(0)/\hat\mu_1(0)$ and $\eta_3(0)=\hat\mu_3(0)/\hat\mu_1(0)$ are the central points of the respective intervals in \eqref{EQ:AtoBC};
\item $\hat\mu_4(0)$ and $\hat\mu_5(0)$ may be selected as the average values of the corresponding parameters for the climate of the macro-area where the plant is located, which are usually available. For example, for the Italian regions considered in the experimental section of this paper such values are estimated in~\cite{BIB:SPENA}.
\end{itemize}
Given their respective definitions, for the N6 model it can be assumed that $\hat\mu_6(0)=\hat\mu_2(0)\hat\mu_4(0)$, while for the L model a natural choice (see \eqref{EQ:LAMBDA}) is given by
$$
\setcounter{MaxMatrixCols}{20}
\begin{array}{rcl}
\theta(\mu) & = & 
 { \left[\begin{array}{ccccc}
								\hat\mu_1(0) 					& 
								\hat\mu_1(0)\hat\mu_4(0) 						& 
								\hat\mu_1(0)\hat\mu_5(0) 						&
								\hat\mu_2(0) 					&
								2\hat\mu_2(0)\hat\mu_4(0) 				
								
					   \end{array}\right.} \\
					   & & {\left.\begin{array}{ccc}
					   			\hat\mu_2(0)\hat\mu_4(0)^2+2\hat\mu_2(0)\hat\mu_5(0)  &
								2\hat\mu_2(0)\hat\mu_4(0)\hat\mu_5(0) 				&
								\hat\mu_2(0)\hat\mu_5(0)^2					\\
								 \end{array}\right.} \\
					   & & {\left.\begin{array}{ccc}
								\hat\mu_3(0) 						&
								\hat\mu_3(0)\hat\mu_4(0) 						&
								\hat\mu_3(0)\hat\mu_5(0)   
								\end{array}\right]^T.}											
\end{array}
$$

\section{Forecasting}\label{sec:forecasting} In this section, the performance of models N5, N6 and L, proposed in Section \ref{SEC:JOIN_MODEL} is evaluated on the widely used Day-Ahead (DA) and Hour-Ahead (HA) forecasts \cite{BIB:STATEOFART}. In order to suitably define them in our context, let $d$ and $k$ be the generic day and time instant, respectively, in which a forecast is supposed to be computed and submitted. For a given time instant $j\geq k$, let $\hat W(j|k)=\{\hat N(j|k),\hat T(j|k)\}$ denote the weather forecast \{cloud cover, temperature\} relative to time $j$ available at time $k$. We denote by $\hat P(j|k;q)$ the prediction of generated power for time instant $j$, computed at time $k$ using the parameter vector estimate available at time $q\leq k$ according to the appropriate model, i.e.,
\begin{equation}\label{EQ:PHAT}
\hat P(j|k;q)=\left\{\begin{array}{ll}
\varphi^T(I^0(j),\hat T(j|k), \hat N(j|k))\cdot \theta(\hat\mu(q)) & {\rm (N5,~N6)} \\
\varphi^T(I^0(j),\hat T(j|k), \hat N(j|k))\cdot\hat\theta(q) & {\rm (L)} .
\end{array}\right.
\end{equation}
\subsection{Day-Ahead forecast}
\begin{figure}[htbp!]
	\centering
	\psfrag{k}{$j$}
    \psfrag{t}{~}
	\psfrag{kd}{${\cal K}_{d+1}$}
	\psfrag{dd}{$d+1$}
	\psfrag{d}{$d$}
	\psfrag{ddd}{$d-1$}
	\psfrag{pp}{\textcolor{blue}{$\hat {\mathcal P}^{DA}(d)$}}
	\psfrag{6}{\small{06:00}}
	\psfrag{n}{$k_0^d$ (present)}
	\psfrag{h}{$\overline k_{d-1}$}
	\includegraphics[width=.8\textwidth]{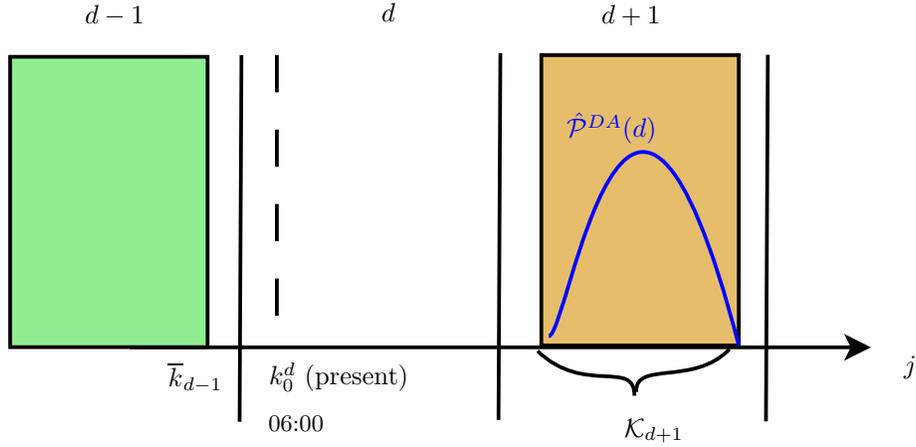}
	\caption{Day-ahead forecasting. Note that ${\cal K}_{d+1}$ does not cover a complete day period since $\overline{k}_{d+1}$ refers to the last sample pertaining to light hours in day $d+1$.}
	\label{FIG:DA}
\end{figure}

The day-ahead forecast is usually submitted at 6 am on the day before each operating day, which begins at midnight on the day of submission, and covers all
24 hours of that operating day. Therefore, the day ahead forecast is provided 19 to 42 hours prior to the operating time. In this respect, we recall that the vast majority of conventional generation is scheduled in the DA market. For a given day $d$, the DA forecast is submitted at time instant $k_0^d$ corresponding, e.g., to $6$ am and consists of the time series given by
\begin{equation}\label{EQ:DAFOR}
\hat {\mathcal P}^{DA}(d)=\left\{\hat P(j|k_0^d;\overline{k}_{d-1}), ~~j\in{\cal K}_{d+1} \right\}.
\end{equation}
Such a forecast is computed on the basis of the last parameter estimate $\hat\mu(\overline{k}_{d-1})$ [$\hat\theta(\overline{k}_{d-1})$] available the day before (see Figure~\ref{FIG:DA} for the meaning of $k_0^d$ and the forecast horizon).
\subsection{Hour-Ahead forecast}
\begin{figure}[htbp!]
	\centering
	\psfrag{k}{$j$}
    \psfrag{t}{\small{TOD}}
	\psfrag{kd}{${\cal K}_{HA}(k)$}
	\psfrag{dd}{$d+1$}
	\psfrag{d}{$d$}
	\psfrag{ddd}{$d-1$}
	\psfrag{pp}{\textcolor{blue}{$\hat {\mathcal P}^{HA}(k)$}}
	\psfrag{6}{\small{06:00}}
	\psfrag{n}{$k$}
	\psfrag{h}{$\overline k_{d-1}$}
	\includegraphics[width=.5\textwidth]{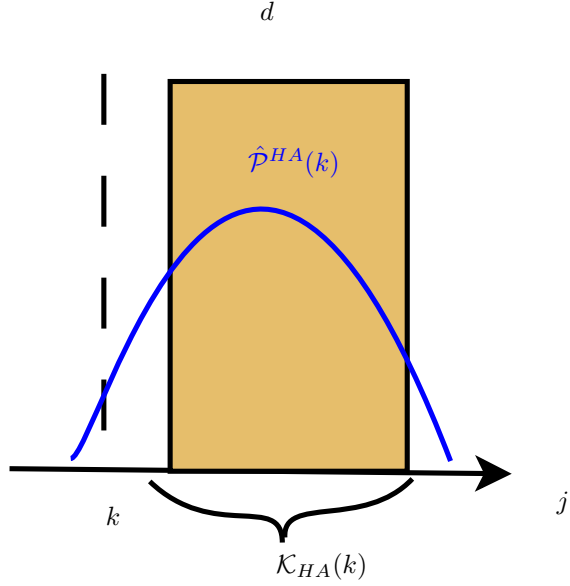}
	\caption{Hour-ahead forecasting}
	\label{FIG:HA}
\end{figure}

The hour-ahead forecast is usually submitted 105 minutes prior to each operating hour and provides an advisory forecast for the 7 hours of light (or the remaining ones, if less) of the same day after the operating hour. Therefore, such a forecast, assuming it is computed at the time index $k$ just before submission, is given by the time series
\begin{equation}\label{EQ:HAFOR}
\hat {\mathcal P}^{HA}(k)=\left\{\hat P(j|k;k), ~~ j\in{\cal K}_{HA}(k)\right\}
\end{equation}
where ${\cal K}_{HA}(k)$ denotes the set of time indices pertaining to the 7-or-less-hour horizon starting at the beginning of the relevant operating hour (see Figure \ref{FIG:HA}). Note that $\hat\mu(k)$ [$\hat\theta(k)$] represents the most recent parameter estimate available.

\section{Performance evaluation}
\label{SEC:PEVA}
In this section we introduce the performance assessment indices that will be used to evaluate the fitness of models to data and the efficacy of the proposed methods in the forecasting problems introduced in the previous section. It is worth to remark that a single estimated model is used to provide both DA and HA forecasts.

\subsection{Error measures}
For the sake of simplicity, a generic definition of the performance indices that will be used is given here. Details on how such indices are computed for specific problems (e.g., DA or HA forecasting) will be provided in the sequel.

Let us consider a data set $\cal P$ 
\begin{equation}
{\cal P} = \left\{\left( P^m(j), \, \hat{P}(j) \right)~~~j\in{\cal K}\right\}
\end{equation}
where $\cal K$ is a set of time indices of cardinality $K$. The set ${\cal P}$ is composed of pairs  $ \left( P^m(j), \, \hat{P}(j) \right)$ with $P^m(j)>0,~\hat P(j)>0$, where $\hat P(j)$ represents the forecasted power and $P^m(j)$ the corresponding measured value. We consider the following standard error measures pertaining to ${\cal P}$:
\begin{itemize}
	\item Root-Mean-Square Error (RMSE) 
		\begin{equation}\label{eq:RMSE}
			RMSE = \sqrt{\frac{1}{K}\sum_{j\in{\cal K}} \left( P^m(j) - \hat{P}(j) \right)^2 }
		\end{equation}
		
	\item Mean-Bias-Error (MBE)
		\begin{equation}\label{eq:MBE}
			MBE = \frac{1}{K} \sum_{j\in{\cal K}} \left( P^m(j) - \hat{P}(j) \right)
		\end{equation}
		
	\item Mean-Absolute-Percentage-Error (MAPE)
		\begin{equation}\label{eq:MAPE}
			MAPE = \frac{1}{K} \sum_{j\in{\cal K}}\abs*{\frac{P^m(j) - \hat{P}(j)}{P^m(j)}} \cdot 100
		\end{equation} 
		
	\item Determination coefficient
		\begin{equation}\label{eq:R}
			R^2 = 1-\frac{\sum_{j\in{\cal K}} \left( P^m(j) - \hat{P}(j) \right)^2 }{\sum_{j\in{\cal K}} \left( P^m(j) - \overline{P} \right)^2 }
		\end{equation}
		where $\overline{P} = \frac{1}{K}\sum_{j\in{\cal K}} P^m(j)$ is the mean of the measured power over the data set. Notice that here it is implicitly assumed that
		$\sum_{j\in{\cal K}} \left( P^m(j) - \hat{P}(j) \right)^2 \leq \sum_{j\in{\cal K}} \Bigl( P^m(j) - \overline{P} \Bigr)^2$ 
		and in this case $1-R^2$ is the so-called unexplained square error. 
\item Normalized RMSE (NRMSE)
			\begin{equation}\label{eq:NRMSE}
			NRMSE = \sqrt{\frac{\sum_{j\in{\cal K}} \left( P^m(j) - \hat{P}(j) \right)^2 }{\sum_{j\in{\cal K}} \left( P^m(j) - \overline{P} \right)^2 }} = \sqrt{1-R^2}.
			\end{equation}
\end{itemize}
In addition to the above standard statistical indices, we also consider the following two error measures, which are of practical interest for network operation, as they are referred to the nominal plant power $P_{nom}$:
\begin{itemize}
	\item Normalized RMSE w.r.t. nominal power ($RMSE_{NP}$)
		\begin{equation}\label{eq:RMSENP}
			RMSE_{NP} = \frac{RMSE}{P_{nom}}
		\end{equation}
	\item Normalized MAPE w.r.t. nominal power ($MAPE_{NP}$)
	\begin{equation}\label{eq:MAPENP}
			MAPE_{NP} = \frac{1}{K} \sum_{j\in{\cal K}}\abs*{\frac{P^m(j) - \hat{P}(j)}{P_{nom}}} \cdot 100 .
		\end{equation} 
\end{itemize}


\subsection{One-day-ahead naive predictor}
As an additional evaluation tool, the performance indices achieved using the proposed approach will be compared to those obtained using the so-called One-Day-ahead Naive Predictor~(ODNP)\mbox{, i.e.,}
\begin{equation}
	\hat{P}^{ODNP}(j) = P^m_{d-1}(j),
\end{equation}
where $P^m_{d-1}(j)$ denotes the measure of generated power recorded during the day before at the same time of day.

\section{Validation on simulated data}
\label{SEC:SIM_RES}
Before assessing the performance of the proposed approach on measured data sets, in this section we present preliminary tests on simulated noisy data in order to highlight the properties of the proposed approach. Nominal weather data (i.e., cloud cover $N(k)$ and temperature $T(k)$) have been generated using an empirical procedure which reproduces a realistic cloud cover and temperature seasonal trend in Italy. The corresponding generated power time series $P^m(k)$ is then computed according to models~\eqref{EQ:PVUSA} and~\eqref{EQ:CCF} using the values of the physical parameters:
\begin{equation}
\label{EQ:PAR_RV}
\begin{array}{l}
	\mu_1 = 0.92,~\mu_2 = \num{-1.237e-4},~\mu_3 = \num{-2.99e-3}, \\
	\mu_4 = -0.3,~\mu_5 = -0.25,~\psi = \SI{27}{\degree},~\zeta = \SI{0}{\degree}. 
\end{array}
\end{equation}
The plant nominal power $P_{nom}$ is assumed equal to \SI{920}{\kilo\watt}.
Generated data are assumed to span the period January 1st (day $\underline d=1$) to December 31st (day $\overline d=365$) and are sampled with a time step  $\tau_s=15$ minutes. The overall data set is then given by
\[
{\cal D}=\left\{D(k)=\{P^m(k),T(k),N(k)\},~~k\in{\cal K}\right\},
\]
where the set of time indices $\cal K$ spans the time horizon from the starting day $\underline d$ to the final day $\overline d$. As previously stated, only the indices $k$ corresponding to hours of light (i.e., for which the theoretical clear-sky irradiance is nonzero) were considered. A model estimation step is therefore performed at each instant $k$ using the data sample $D(k)$.

\subsection{Scenarios}
Five different simulation scenarios (i)$-$(v) have been considered. Within each scenario, different noise and data processing conditions identified by a Setup ID (SID) have been simulated. The scenarios are arranged as follows:
\begin{enumerate}
\item[(i)] Neither noise nor quantization is added. Raw data (SID $0$) and data averaged over \SI{1}{\hour} (SID~$1$) (see Table~\ref{TAB:NOISE_CL}) were considered. Averaging is introduced in order to mimic the typical setting where hourly updates of weather and generation data are provided by the meteorological service and the DSO, respectively.
\begin{table}
\centering
\setlength\tabcolsep{4pt}
\begin{minipage}{0.48\textwidth}
	\centering
	\begin{tabular}{c|c}
		\toprule
		SID	& Description	\\
		\midrule	
		$0$	&  \TwoRowCell{No noise added, raw data sampled at \SI{15}{\minute}}	\\	
		$1$	&  \TwoRowCell{No noise added, data averaged over \SI{1}{\hour}} 		\\	
		\bottomrule
	\end{tabular}
	\caption{Setup descriptions for scenario (i).}
	\label{TAB:NOISE_CL}
\end{minipage}
\hfill
\begin{minipage}{0.48\textwidth}
	\centering
	\begin{tabular}{c|ccccccccccc}
		\toprule
		SID 				& $2$  	& $3$ 		& $4$  		& $5$  	\\ 
		\midrule
		$3\sigma_N$ 		& $0.0$	& $0.1$		& $0.5$		& $1.0$	\\ 
		$\sigma_{N,tot}$ 	&$0.289$& $0.291$	& $0.333$ 	& $0.441$ \\ 
		\bottomrule
	\end{tabular}
	\caption{Setup description for scenario (ii).}
	\label{TAB:NOISE_N}
\end{minipage}
\end{table}


\item[(ii)] A zero-mean Gaussian white noise with variance ${\sigma^2_N}$ is added to $N(k)$, $N(k)$ is then quantized so that $N(k) \in \left\lbrace 0, \, 0.1, \, 0.2, \, \ldots, \, 1.0 \right\rbrace$, resulting in a noise variance $\sigma^2_{N,tot}$ (SID~$2-5$ in Table~\ref{TAB:NOISE_N}). Data is finally averaged over \SI{1}{\hour}.

\item[(iii)] A zero-mean Gaussian white noise with variance $\sigma_P^2$ is added to $P^m(k)$, $N(k)$ is quantized and data are averaged over \SI{1}{\hour} (SID~$6-8$, Table~\ref{TAB:NOISE_PTN}).
\item[(iv)]  A zero-mean Gaussian white noise with variance $\sigma_T^2$ is added to $T(k)$, $N(k)$ is quantized and data are averaged over \SI{1}{\hour} (SID~$9-11$, Table~\ref{TAB:NOISE_PTN}).
\item[(v)] A zero-mean Gaussian white noise is added to all the regression variables, $N(k)$ is quantized and data are averaged over \SI{1}{\hour} (SID~$12$, Table~\ref{TAB:NOISE_PTN}).

\begin{table}
\centering
\begin{tabular}{l|ccccccc}
\toprule
SID 									& $6$	& $7$	& $8$  	& $9$  	& $10$  & $11$ 	& $12$ \\ 
\midrule
$3\sigma_N$ 							& $0.0$	& $0.0$	& $0.0$	& $0.0$	& $0.0$	& $0.0$	& $0.3$	\\ 
$3\sigma_P \, (\si{\kilo\watt})$		& $10$ 	& $50$ 	& $100$	& $0$ 	& $0$	& $0$ 	& $50$ 	\\ 
$3\sigma_T \, (\si{\degreeCelsius})$	& $0$	& $0$	& $0$	& $1$	& $3$	& $5$	& $3$	\\
\bottomrule
\end{tabular}
\caption{Setup description for scenarios (iii), (iv) and (v).}
\label{TAB:NOISE_PTN}
\end{table}
\end{enumerate}
Notice that here data averaged over \SI{1}{\hour} means that for each hour $h$ there is only one sample computed as mean of the data acquired during $h$.

The performance of the proposed method on the scenarios introduced above has been evaluated with reference to the day-ahead (DA) forecast introduced in Section~\ref{sec:forecasting}. For computing the forecasts, weather data from the data set $\cal D$ have been employed. In particular, at each time instant $k_0^d$ corresponding to 6 am on each day, we evaluate the DA forecast $\hat P^{DA}(d)$ according to \eqref{EQ:PHAT}-\eqref{EQ:DAFOR} with
\[
\hat T(j|k_0^d)=T(j),~~\hat N(j|k_0^d)=N(j),~~j\in{\cal K}_d.
\]

The evaluation of the performance indices starts at day $d_0=18$, in order to first ensure a rough degree of adaptation of the model parameters.
The following performance indices were considered:
\begin{itemize}
\item $RMSE_d$: the RMSE relative to the forecast pertaining to each day $d$, i.e., \eqref{eq:RMSE} evaluated for $\hat P(j)=\hat P(j|k_0^d;\overline k_{d-1})$, ${\cal K}={\cal K}_d$;
\item RMSE, MBE, MAPE, and determination coefficient $R^2$, evaluated over the whole simulation, i.e., over the union of all DA forecasts from day $d_0$ to day $\overline d$.
\end{itemize}

\enlargethispage{\baselineskip}
For each SID, $10$ simulations were performed, in which $P^m(k)$, $T(k)$ and $N(k)$ were obtained by using different noise realizations. The average of performance indices over all the simulations were considered.

In each scenario, the initial values $\hat\mu_1(0)$, $\hat\mu_2(0)$, $\hat\mu_3(0)$, $\hat\mu_4(0)$ and $\hat\mu_5(0)$ are chosen as $75\%$ of the real values in~\eqref{EQ:PAR_RV}, while $\psi$ and $\zeta$ are equal to their nominal values. $l(0)$ has been initialized to $0.01$ for both RLS algorithms and EKF, $r$ has been initialized to $10^4$, which is a conservative choice in view of the variance of the noise added to power measurements.

\subsection{Results}
\begin{figure*}[htbp!]
\parRP
\centering
\subfloat[][\emph{Scenario (i) - SID $0$, noise-free data}]{%
	\includegraphics[width=1\textwidth]{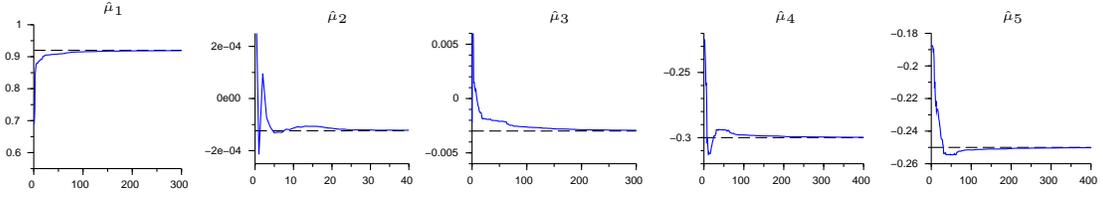}\label{FIG:SIM_PAR_CLEAN}} \\
\subfloat[][\emph{Scenario (i) - SID $1$, noise-free averaged data}]{%
	\includegraphics[width=1\textwidth]{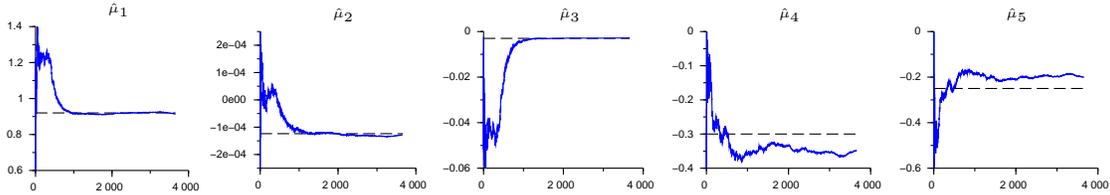}\label{FIG:SIM_PAR_NONOISE}} \\
\caption{N5 model parameter estimation vs. time (iterations), scenario (i). Black dashed lines are the nominal values of each parameter as reported in~\eqref{EQ:PAR_RV}.}
\label{FIG:SIM_RP_PARAMETER_1}
\end{figure*}

Figure~\ref{FIG:SIM_RP_PARAMETER_1} shows one instance of the parameter estimate evolution for the N5 model for SID~$0$ and SID~$1$. Using noise-free raw data (SID 0) the parameters converge exactly to the real values (Figure~\ref{FIG:SIM_PAR_CLEAN}). In case of SID~$1$ the parameters converge as well with slight deviations of $\hat\mu_4$ and $\hat\mu_5$ from the nominal values (Figure~\ref{FIG:SIM_PAR_NONOISE}). Note that in this case, the parameters take considerably longer to converge.

\begin{figure*}[htbp!]
\parRP
\centering
\subfloat[][\emph{Scenario (ii) - SID $5$, noise added only to $N$.}]{%
	\includegraphics[width=1\textwidth]{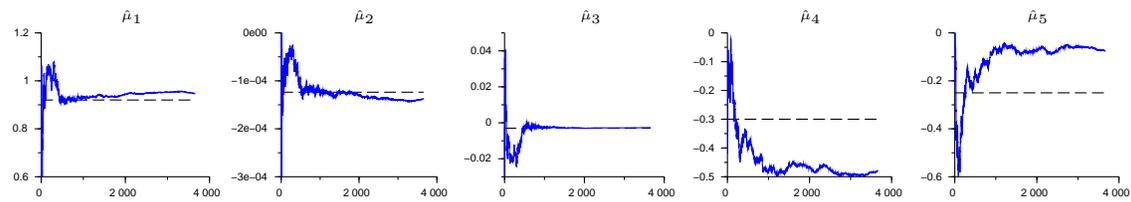}\label{FIG:SIM_PAR_NOISE_N}} \\	
\subfloat[][\emph{Scenario (iii) - SID $8$, noise added only to $P$.}]{%
	\includegraphics[width=1\textwidth]{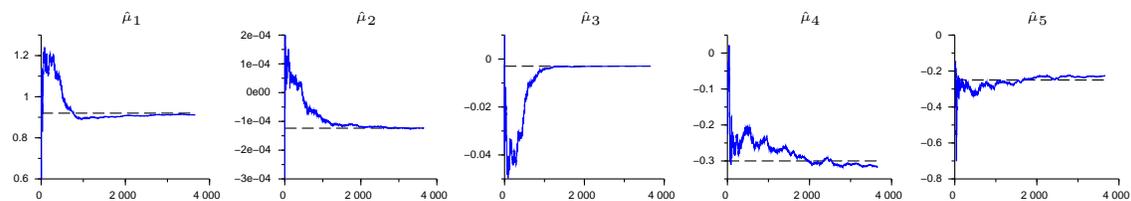}\label{FIG:SIM_PAR_NOISE_P}} \\	
\subfloat[][\emph{Scenario (iv) - SID $11$, noise added only to $T$.}]{%
	\includegraphics[width=1\textwidth]{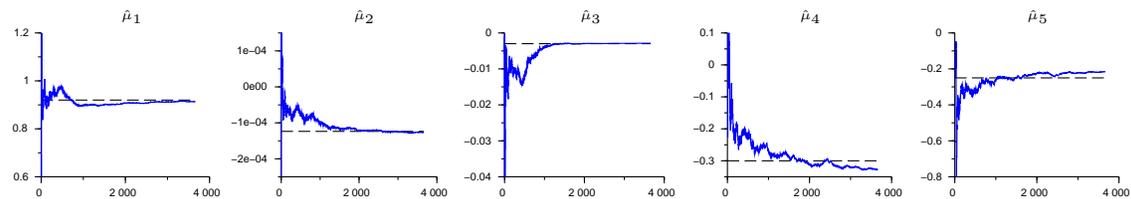}\label{FIG:SIM_PAR_NOISE_T}} \\	
\subfloat[][\emph{Scenario (v) - SID $12$, noise added to $N$, $P$ and $T$.}]{%
	\includegraphics[width=1\textwidth]{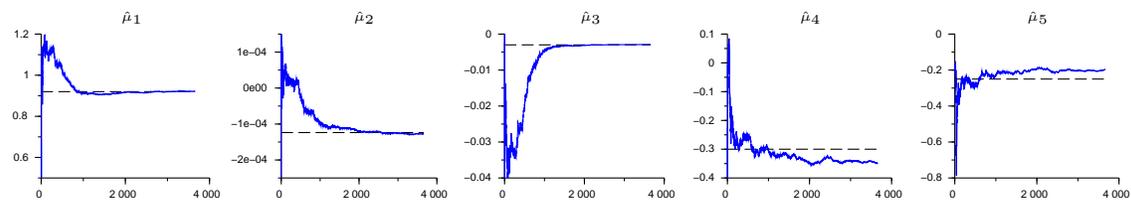}\label{FIG:SIM_PAR_NOISE_NPT}} \\	
\caption{N5 model parameter estimation (blue line) vs. time (iterations) and using data from scenario (ii) to scenario (v). Black dashed lines are the nominal values of each parameter as reported in~\eqref{EQ:PAR_RV}.}
\label{FIG:SIM_RP_PARAMETER_2}
\end{figure*}

Figures~\ref{FIG:SIM_PAR_NOISE_N}, \ref{FIG:SIM_PAR_NOISE_P} and~\ref{FIG:SIM_PAR_NOISE_T} show three examples of parameter estimate evolution for the N5 model under the largest noise variance on $N$, $P$ and $T$, respectively, Figure~\ref{FIG:SIM_PAR_NOISE_NPT} shows the same plot when noise is added to all the regression variables. Parameters $\hat\mu_4$ and $\hat\mu_5$ are apparently quite sensitive to the noise added to $N$. Indeed, under the conditions described by the SID~$5$, for which $\sigma_{N,tot} = 0.441$, $\hat\mu_4$ is underestimated and $\hat\mu_5$ is overestimated (Figures~\ref{FIG:SIM_PAR_NOISE_N}). All the other parameters always tend to converge to their real values.

\begin{figure*}[htbp!]
\rmse
\centering
\vspace*{-2cm}
\subfloat[][\emph{SID $5$, noise added only to $N$.}]{%
	\includegraphics[width=1.2\textwidth]{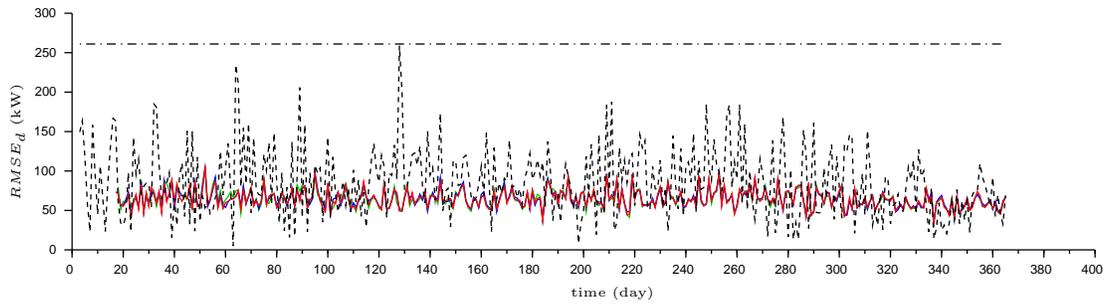}\label{FIG:RMSE_N}} \\
\vspace*{-2.5mm}
\subfloat[][\emph{SID $8$, noise added only to $P$.}]{%
	\includegraphics[width=1.2\textwidth]{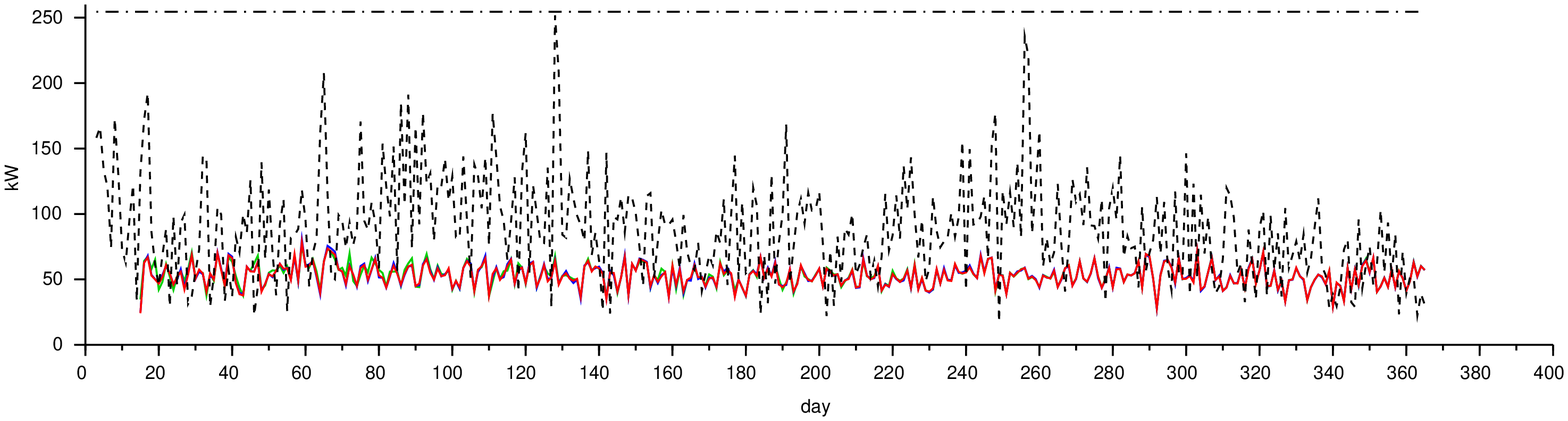}\label{FIG:RMSE_P}} \\
\vspace*{-2.5mm}
\subfloat[][\emph{SID $11$, noise added only to $T$.}]{%
	\includegraphics[width=1.2\textwidth]{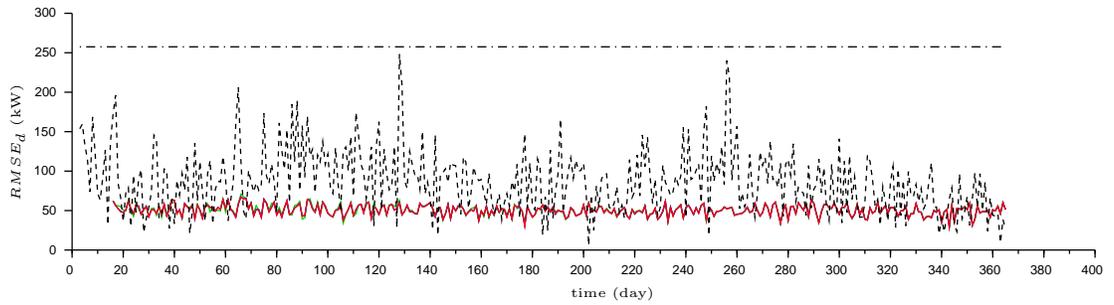}\label{FIG:RMSE_T}} \\
\vspace*{-2.5mm}
\subfloat[][\emph{SID $12$, noise added to $N$, $P$ and $T$.}]{%
	\includegraphics[width=1.2\textwidth]{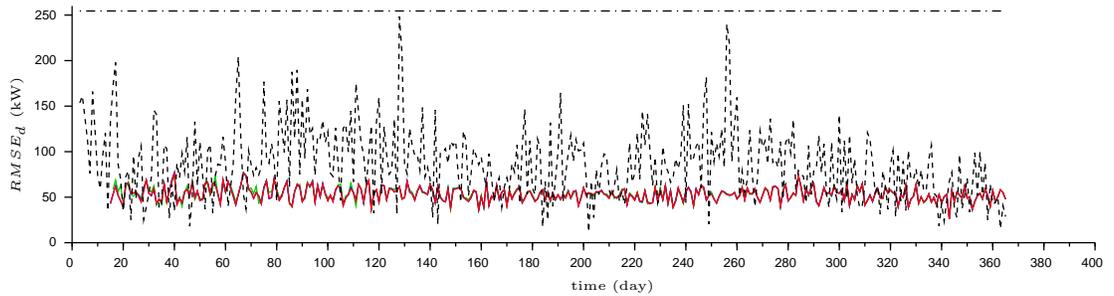}\label{FIG:RMSE_NPT}} \\
\caption{Comparison between the standard deviation of the simulated power (dashed dotted line), the $RMSE_d$ computed using the ODNP (dashed line), L model (black line), N5 model (blue line) and N6 model (red line).}
\label{FIG:SIM_RMSE}
\end{figure*}

Figure \ref{FIG:SIM_RMSE} depicts the evolution of the daily RMSE ($RMSE_d$) for increasing SID over the whole simulation (365 days), relative to L, N5, N6 models and to the ODNP. Such error is compared with the standard deviation of simulated power data. In view of the fact that parameter convergence behaves differently for different SIDs and in order to have a consistent comparison, the performance indices for all cases were computed starting from day $15$.
Results show that the proposed approach is more sensitive to noise added to $N$, than to noise added to $P$ or $T$. The $RMSE_d$ computed with SID~$5$ (Figure~\ref{FIG:RMSE_N}) is considerably higher than the $RMSE_d$ computed with SID~$8$ and~$11$ (Figures~\ref{FIG:RMSE_P} and~\ref{FIG:RMSE_T}), which are really similar to each other. Clearly, under any of the tested noise conditions, the $RMSE_d$ achieved by all three models is by far lower than both the $RMSE_d$ obtained using the ODNP and the standard deviation of the simulated power.

\begin{figure}[htbp!]
	\psfrag{rmse}[b]{\tiny $RMSE (\si{\kilo\watt})$}
	\psfrag{mape}[b]{\tiny $MAPE$}
	\psfrag{mbe}[b]{\tiny $MBE (\si{\kilo\watt})$}
	\psfrag{r2}[b]{\tiny $R^2$}
	\psfrag{dataID}{\tiny SID}
	\centering
	\subfloat[][\emph{$RMSE$ and $MAPE$.}]{%
		\includegraphics[width=1\columnwidth]{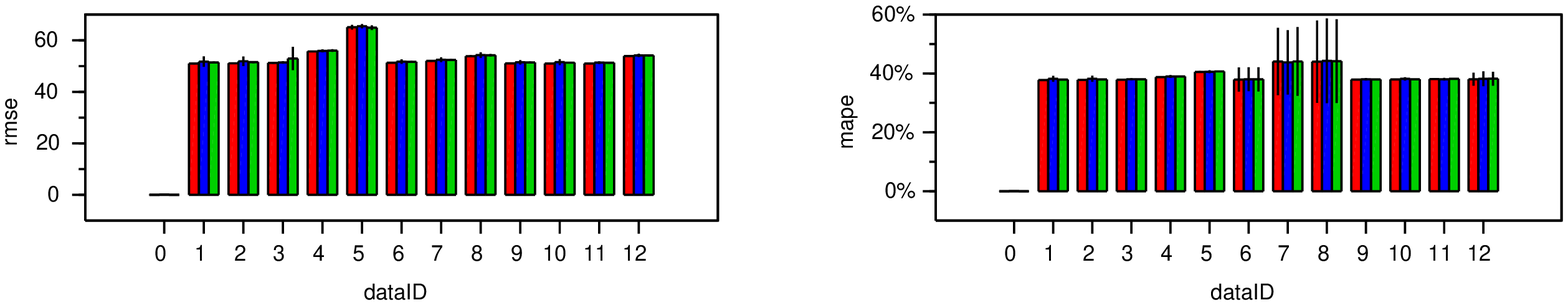}\label{FIG:RMSE_MAPE_NOANGLE}} \\
	\subfloat[][\emph{$MBE$ and $R^2$.}]{%
		\includegraphics[width=1\columnwidth]{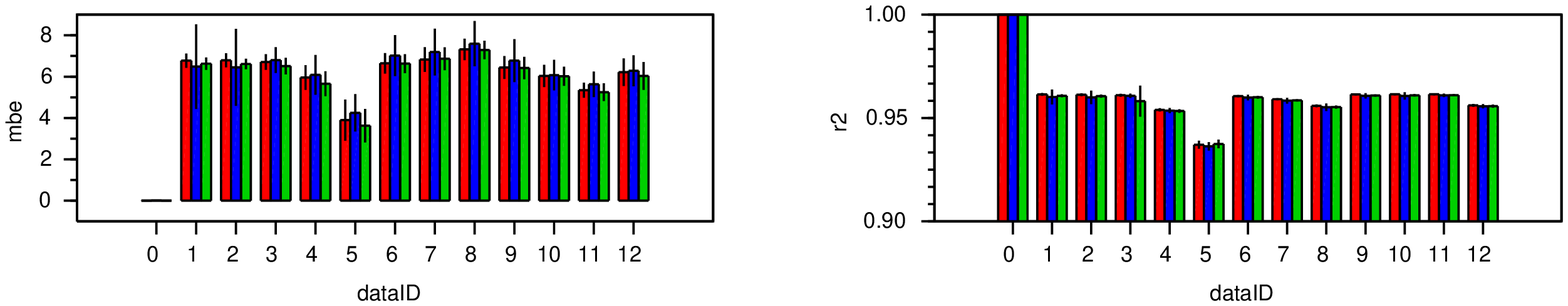}\label{FIG:MBE_R2_NOANGLE}} \\
	\caption{Performance indices vs. noise computed using simulated data. Red bars represent the N5 model, blue bars the N6 model, and green bars the L model. }
	\label{FIG:SIM_EC_NOANGLE}
\end{figure}

Performance indices RMSE, MAPE, MBE, and $R^2$ over the whole simulation are reported and compared for each SID and each model in Figure~\ref{FIG:SIM_EC_NOANGLE}. 
$RMSE$ and $MAPE$ slowly increase and $R^2$ slowly decreases with $\sigma_N$ (SID~$2-5$) and $\sigma_P$ (SID~$6-8$), while they are almost constant when $\sigma_T$ (SID~$9-11$) increases. Finally, note that $MBE$ grows when $\sigma_P$ is increased and that it decreases both when $\sigma_T$ is increased and when $\sigma_N$ grows.

%
\begin{figure}[htbp!]
	\power
	\centering
	\includegraphics[width=\textwidth]{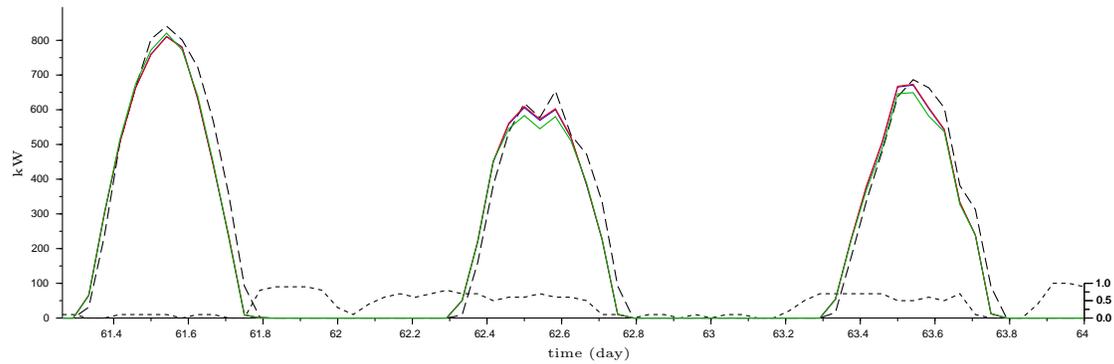}
	\caption{Scenario (v) - SID~$12$, comparison between the simulated power (long dashed line),  N5 model DA forecast (red line), N6 model DA forecast (blue line, which almost overlaps with the red line) and L model DA forecast (green line). Short dashed line represents the CCI (scale to the right).}
	\label{FIG:SIM_FORECAST}
\end{figure}

In Figure \ref{FIG:SIM_FORECAST}, the power forecasts provided by models L, N5 and N6 during a simulation run with SID~$12$ are qualitatively compared to measured data during three consecutive days. The picture refers to typical examples of partially cloudy days.
\clearpage
\section{Validation on experimental data}
\label{SEC:EXP_RES}
\subsection{Experiment set up}
The proposed procedure has been validated using data from a photovoltaic plant of nominal power $P_{nom} = \SI{920}{\kilo\watt}$p located in Sardinia (Italy). Available data consist of two datasets spanning February 2nd (33rd day of year) to May 1st (122nd day of year):
\begin{itemize}
	\item ${\cal D}_1$: a set of data provided by the private producer running the plant, which collects hourly samples of averaged measured power and one day-ahead forecasts of air temperature;
	\item ${\cal D}_2$: a set of weather reports including CCI, evaluated by a meteorological station located \SI{20}{\kilo\metre} away from the plant. This data set does not have a regular sampling time. 
\end{itemize}
${\cal D}_1$ and ${\cal D}_2$ have been preprocessed in order to have a single dataset ${\cal D}$ of hourly data. If there are more then one CCI values in ${\cal D}_2$ for the same hour, an averaged CCI is computed for that hour. On the contrary, if the CCI report in ${\cal D}_2$ is missing for a given hour, then the measured power and the forecast of air temperature in ${\cal D}_1$ for that hour are dropped\footnote{For the sake of completeness, and to further support the claim that one day-ahead temperature forecasts can be used in our procedure in place of actual data, temperature measurements collected at the meteorological station have been compared with the forecasts included in ${\cal D}_1$. The RMSE and MBE computed on the overall data set turned out to be equal to $1.9^\circ C$ and $0.9^\circ C$, respectively.}.
The data set employed is therefore given by the time series 
\[
{\cal D}=\left\{D(k)=\{P^m(k),T(k),N(k)\},~~k\in{\cal K}\right\},
\]
where the set of time indices $\cal K$ spans, with a sampling time $\tau_s = \SI{1}{\hour}$, from the starting day $\underline d=33$ to the final day $\overline d=122$. Also in this case, only the indices $k$ corresponding to hours of light were considered. Model parameter updates are computed hourly using the data sample $D(k)$.

The initial values of the parameters have been chosen as follows \footnote{Concerning the orientation angles $\psi$ and $\zeta$, which were not available, their values have been guessed by visually comparing the generation curve during a clear-sky day with the clear-sky irradiance model.}:
\[
\begin{array}{l}
	\hat\mu_1(0) = P_{nom}/1000,~\hat\mu_2(0) = \num{-1.34e-4}\cdot \hat\mu_1(0),~\hat\mu_3(0) = \num{-3.25e-3}\cdot \hat\mu_1(0), \\
	\hat\mu_4(0) = 0.784,~\hat\mu_5(0) = -1.344,~\psi = \SI{27}{\degree},~\zeta = \SI{12}{\degree}, \\
	l(0)=10,~r=10^4.
\end{array}
\]

\subsection{Performance evaluation}
\label{SEC:EVALUATION}
The performance of the proposed method on the data set introduced above has been evaluated with reference to the two standard forecast types introduced in Section~\ref{sec:forecasting}, i.e.,
\begin{itemize}
\item {\sl DA forecast}.
In this case, the forecasts and the performance indices $RMSE_d$, RMSE, MAPE, MBE and $R^2$ were computed using the same modality as in the previous section.
\item {\sl HA forecast}.
The HA forecasts are computed according to \eqref{EQ:PHAT}-\eqref{EQ:HAFOR} based on the data set $\cal D$ using
\[
\hat T(j|k)=T(j),~~\hat N(j|k)=N(j),~~j>k.
\]
In order to evaluate the performance, a single 7-hour HA forecast per day has been considered, and the error measures RMSE, MAPE, MBE and $R^2$ have been computed over the the union of all such forecasts.
\end{itemize}

The performance of the proposed method has been compared with that obtained using the following approaches.
\begin{itemize}
	\item ODNP (for DA forecasts only, clearly not suitable in the HA case). 
	\item Autoregressive prediction models using only measurements of generated power and only cloud cover data, respectively. These methods represent a more realistic benchmark with respect to ODNP and are applicable to both DA and HA. To this end, the following two models have been tested:	
	\begin{itemize}
		\item PVGM: autoregressive (AR) model of generated power with regression horizon equal to 12 hours. The parameter vector is given by $a = \left[ a_1 \ldots a_{12} \right]^T$ and the regressor at time $j$ is expressed by $\psi(j) = \left[ P(j-1) \ldots P(j-12) \right]^T$. Then the prediction of generated power for time instant $j$, computed at time $k$ using parameters available at time $q \leq k$ is:
		\[
			\hat{P}(j|k,q) = \hat{a}(q)^T \psi(j),
		\]
		where: $\hat{a}(q)$ is the estimate of $a$ available at time $q$, and for $i=1\,\ldots\,12$, the $i$-th term $P(j-i)$ of $\psi(j)$ is equal to $P^m(j-i)$ if $j-i \leq k$, and equal to $\hat{P}(j-i|k,q)$ otherwise.
		 
		\item CCD: autoregressive with exogenous input (ARX) model of generated power with regression horizon equal to 2 hours. The inputs are cloud cover data and the clear-sky model~\eqref{EQ:ICS0}\footnote{Cloud cover index alone cannot possibly provide enough information.}. The parameter vector is given by $b = \left[ b_1 \ldots b_{6} \right]^T$ and the regressor at time $j$ is expressed by 
		$\xi(j) = \left[ I^0(j) \ldots I^0(j-2)~~N(j) \ldots N(j-2)  \right]^T$. Then the prediction of generated power $\hat{P}(j|k,q)$ is:
		\[
			\hat{P}(j|k,q) = \hat{b}(q)^T \xi(j),
		\]
		where: $\hat{b}(q)$ is the estimate of $b$ available at time $q$, and for $i=0,\,1,\,2$, the $i$-th term $N(j-i)$ of $\xi(j)$ is equal to $N^m(j-i)$ if $j-i \leq k$, and equal to $\hat{N}(j-i|k)$ otherwise. 
	\end{itemize}
	The parameters of the two models above have been estimated in the same recursive framework as the proposed parametric model by means of Linear Least Squares.
	
	\item ANN-based approach, using the MultiLayer Perceptron (MLP) architecture. 
	In this case we are looking for a fair comparison in terms of model complexity, therefore two architectures with the same number of parameters as the N6 and L parametric models, respectively, have been tested using the training algorithms implemented in the MATLAB Neural Network Toolbox. Two architectures have been selected:	
	%
	\begin{itemize}
		\item MLP1: $1$ hidden unit, $6$ parameters,
		\item MLP2: $2$ hidden units, $11$ parameters.
	\end{itemize}

	Note that no such comparison is possible with the N5 model since a 5-parameter MLP cannot be synthesized. The networks have been trained in the same recursive framework as the proposed parametric model, but since a training step for the MLPs cannot be performed for each incoming data sample $D(k)$, the algorithm has been modified in order to build suitable learning sets for network training. In particular, a new learning set is constructed every $75$ samples.
\end{itemize}

\subsection{Results and discussion}

\begin{figure*}[htbp!]
	\parRP
	\centering
	\includegraphics[width=1\textwidth]{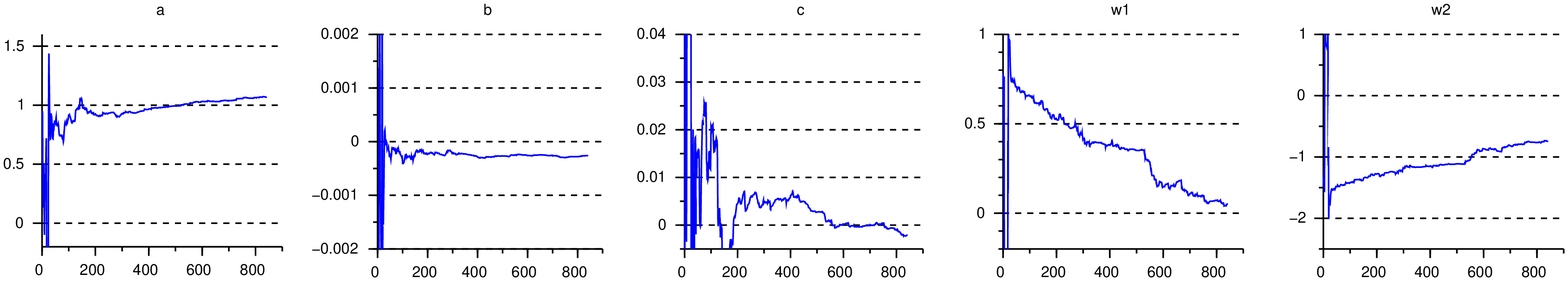} \\
	\caption{N5 model parameter estimation. Parameter estimates vs. time (iterations).}
	\label{FIG:RP_PARAMETER}
\end{figure*}
\begin{figure*}[htbp!]
	\parRP
	\centering
	\psfrag{bw1}{\tiny $\hat\mu_6$}
	\includegraphics[width=1\textwidth]{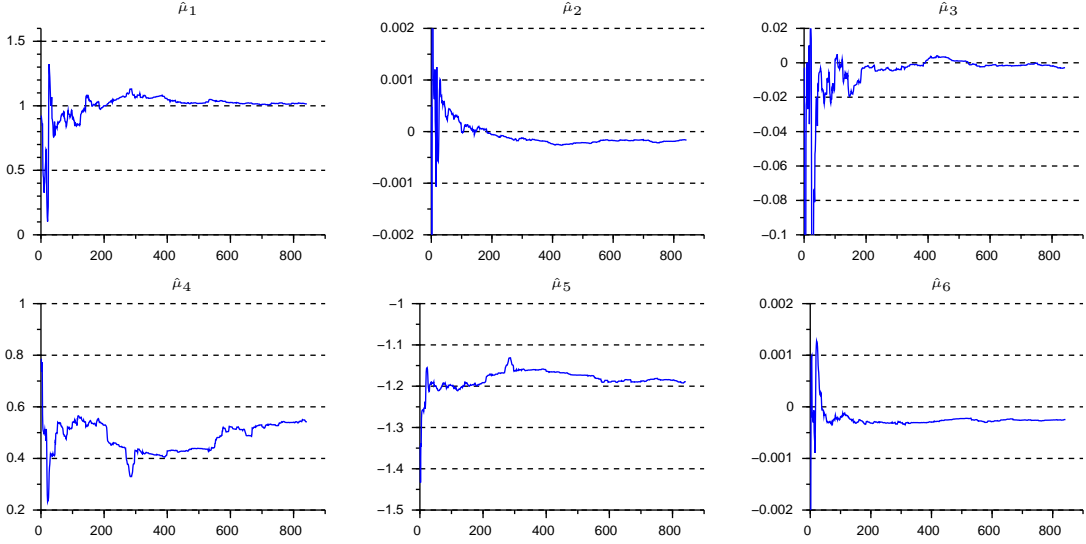} \\
	\caption{N6 model parameter estimation, parameter estimates vs. time (iterations).}
	\label{FIG:RP_PARAMETER6}
\end{figure*}

In Figures~\ref{FIG:RP_PARAMETER} and \ref{FIG:RP_PARAMETER6} the evolution of the parameter estimates are reported for the N5 and N6 models. All the parameters tend to converge for these as well as for the L model, not depicted to save space. The estimated parameter values and their standard deviation at the end of the experiment are shown below:

%
\begin{align}
(N5)
&
\left.
\begin{array}{rcl}
	&\hat\mu=	& 	\left[
					\begin{array}{ccccc}
    					\fnsnum{1.03} & \fnsnum{-2.54e-4} & \fnsnum{-2.12e-4} & \fnsnum{1.29e-1} &	\fnsnum{-8.42e-1} \\				
					\end{array}
					\right]^T,
					\\
	&\sigma_{\hat\mu}= &
					\left[
					\begin{array}{ccccc}
						\fnsnum{1.72e-2} & \fnsnum{1.58e-5}	& \fnsnum{7.30e-4} & \fnsnum{6.02e-2} & \fnsnum{8.00e-2} \\
					\end{array}
					\right]^T,
\end{array}
\right.\label{EQ:N5_PAR}
\end{align}
\begin{align}
(N6)
&
\left.
\begin{array}{rcl}
	&\hat\mu=	&   \left[
					\begin{array}{cccccc}
    					\fnsnum{1.01} & \fnsnum{-1.60e-4} & \fnsnum{-1.75e-3} & \fnsnum{5.21e-1} & \fnsnum{-1.19} & \fnsnum{-2.53e-4} \\				
					\end{array}
					\right]^T,
					\\
	&\sigma_{\hat\mu}= &
					\left[
					\begin{array}{cccccc}
						\fnsnum{8.88e-2} & \fnsnum{1.57e-5}	& \fnsnum{6.24e-4} & \fnsnum{2.52e-2} & \fnsnum{3.82e-3} & \fnsnum{1.36e-5} \\
					\end{array}
					\right]^T,
\end{array}
\right.\label{EQ:N6_PAR}
\end{align}
\begin{align}
(L)
&
\left.
\begin{array}{rcl}
	&\hat\mu=	&   \left[
					\begin{array}{cccccc}
						\fnsnum{1.39} & \fnsnum{-9.54e-1} & \fnsnum{-2.53e-1} & \fnsnum{-3.49e-4} & \fnsnum{6.00e-4} & \fnsnum{-1.79e-3} \\
					\end{array}
					\right. 
					\\
	&			&	\left.	
					\begin{array}{ccccc}
						\fnsnum{3.45e-3} & \fnsnum{-1.75e-3} & \fnsnum{-1.40e-2} & \fnsnum{3.87e-2} & \fnsnum{-2.47e-2} \\					
					\end{array}
					\right]^T,
					\\
	&\sigma_{\hat\mu}= &
					\left[
					\begin{array}{cccccc}
						\fnsnum{1.66e-2} & \fnsnum{9.18e-2} & \fnsnum{9.27e-2} & \fnsnum{2.04e-5} & \fnsnum{2.11e-4} & \fnsnum{5.94e-4} \\
					\end{array}
					\right. 
					\\
	&			&	\left.	
					\begin{array}{ccccc}
						\fnsnum{9.85e-4} & \fnsnum{5.79e-4} & \fnsnum{1.21e-3} & \fnsnum{9.10e-3} & \fnsnum{1.02e-2} \\					
					\end{array}
					\right]^T.
					\\
\end{array}
\right.\label{EQ:L_PAR}
\end{align}

\enlargethispage{-\baselineskip}
Notice that all standard deviations are at least one order of magnitude lower than the respective parameter estimates. The only exception concerns the parameter estimate $\hat\mu_3$ for the N5 model. This model, which is characterized by a minimal number of parameters, apparently does not fully capture the dependency between the generated power and temperature, thus generating persistent parameter drifts. On the contrary, the slightly overparameterized N6 model shows much better convergence properties, as it is also apparent from the comparison of Figures \ref{FIG:RP_PARAMETER} and \ref{FIG:RP_PARAMETER6}. Furthermore, the L model shows higher parameter standard deviations, on average, with respect to N6.

In order to analyze the impact of overparametrization on the consistency of the parameter estimates with the physical model \eqref{EQ:PVUSAK}, we compare the values of $\hat\mu_1$ and the correction terms $\hat\eta_2=\hat\mu_2/\hat\mu_1$ and $\hat\eta_3=\hat\mu_3/\hat\mu_1$ obtained from~\eqref{EQ:N5_PAR}, \eqref{EQ:N6_PAR} and \eqref{EQ:L_PAR}.
The estimates of $\hat{\mu}_1$ provided by models N5 and N6 are slightly higher than the nominal power/irradiance gain $P_{nom}/1000 = 0.92$, while the value obtained using model L is remarkably higher. In Table~\ref{TAB:ETA}, $\hat\eta_2$ and $\hat\eta_3$  are compared with their typical values reported in~\eqref{EQ:AtoBC}. Notice that N6 is the only model for which $\eta_2 \in {\cal S}_2$ and $\eta_3 \in {\cal S}_3$, while for N5 $\hat\eta_3$ is higher than the upper bound of ${\cal S}_3$, which apparently means that $\mu_3$ is overestimated. On the contrary, model L tends to underestimate $\mu_3$.
\begin{table}[htbp!]
	\centering
	\begin{tabular}{ccccc}
	\toprule
	$\hat\eta_i$	& ${\cal S}_i$							  & N5			   & N6			    & L			   \\
	\midrule	
	$\hat\eta_2$	& $\left[\num{-2.5e-4}, \num{-1.9e-5}\right]$ & \num{-2.47e-4} & \num{-1.58e-4} & \num{-2.51e-4} \\
	$\hat\eta_3$	& $\left[\num{-4.8e-3}, \num{-1.7e-3}\right]$ & \num{-2.06e-4} & \num{-1.73e-3}  & \num{-10.1e-3} \\
	\bottomrule
	\end{tabular}
	\caption{Correction terms $\eta_2$ and $\eta_3$ computed using the estimated parameter vectors reported in~\eqref{EQ:N5_PAR},~\eqref{EQ:N6_PAR}~and~\eqref{EQ:L_PAR}.}
	\label{TAB:ETA}
\end{table}

Notice that in model N6 the overparametrization is given only by the introduction of parameter $\mu_6$, defined as the product between $\mu_2$ and $\mu_4$.  From~\eqref{EQ:N6_PAR}, we have that $\hat{\mu}_2 = \num{-1.60e-4}$, $\hat{\mu}_4 = -1.19$, and $\hat{\mu}_6 = \num{-2.53e-4}$. Therefore $\hat{\mu}_2\hat{\mu}_4 = \num{-9.12e-4}$, which is the same order of magnitude as $\hat{\mu}_6$, thus suggesting that the estimates are almost consistent. A similar consistency is not shown by model L, despite its good performance on the DA and HA forecasts, as shown in the sequel.

\begin{figure}[htbp!]
	\centering
	\psfrag{C(N)}{\tiny{$C(N)$}}
	\psfrag{N}{\tiny{$N$}}
	\includegraphics[width=.45\textwidth]{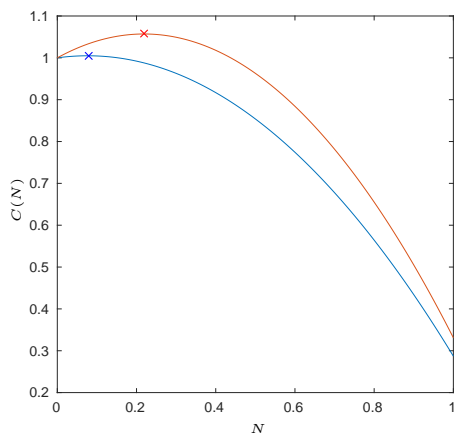}
	\caption{Comparison between the CCF curves obtained using the estimated values of $\mu_4$ and $\mu_5$ provided by models N5 (blue line) and N6 (red line), x-marks denote the maximum points.}
	\label{FIG:ESTIMATED_CCF}
\end{figure}

In order to further evaluate the consistency of models N5 and N6, we analyze the CCF curve $C(N)$ in  \eqref{EQ:KIMURA} obtained using the estimated values of $\mu_4$ and $\mu_5$ provided by such models. In Figure \ref{FIG:ESTIMATED_CCF} we report the plots of the $C(N)$ using the respective values of $\hat{\mu}_4$ and $\hat{\mu}_5$. It turns out that both the estimated CCFs are consistent with the seasonal trend for the mediterranean climate \citep{BIB:SPENA}.

\enlargethispage{-1\baselineskip}
As far as the forecasting performance evaluation is concerned, all error measures were computed over the period starting from day $d_0=57$ in order to guarantee, as in the simulated data case, at least a rough adaptation of the model parameters.

Table \ref{TAB:MLP_CMP_ONLINE} summarizes the performance indices achieved by the proposed parametric models, compared with MLPs, ODNP, PVGM and CCD on both DA and HA forecasts. Since the performance of the MLPs may be quite sensitive to the initial conditions, 30 simulations of MLP1 and MLP2 with the same data set have been performed with suitably generated random initial conditions\footnote{MLP layer weights and biases are initialized according to the Nguyen-Widrow algorithm. This algorithm chooses values in order to distribute the active region of each neuron in the layer randomly but evenly across the layer's input space.}, and mean value/variance of all performance indices over all simulations have been reported. Clearly, all parametric models and MLPs perform largely better than the ODNP and PVGM. On average, MLPs show errors 20\% higher than parametric models. The CCD model yields intermediate indices between our models and MLPs. The performances of the three proposed parametric models are overall comparable. Notice that in network operation, performance indices qualifying forecast accuracy, e.g., RMSE and MAPE, are generally normalized with respect to the nominal plant power. This normalization is useful in view of how the DSO quantifies uncertainty in power generation. Actually, when dealing with hundreds or thousands of distributed units, either for maintenance activities or operational requirements, the DSO takes into account how much each specific unit is expected to produce with respect to its nominal power. Indeed, network operation requires the DSO to know at any time the overall power uncertainty pertaining to the whole distributed generation. Such uncertainty can be easily estimated from the nominal capacity of each plant (known a-priori) and the normalized plant forecast error (computed off-line on historical data and usually with a smaller time resolution, typically, weekly or monthly). For this reason, we consider the normalized errors $RMSE_{NP}$ and $MAPE_{NP}$. The results obtained for the latter indices, which are around  8-10\%, can be considered acceptable for most problems in network operation.
\begin{table}[htbp!]
	\hspace*{-1cm}
	\begin{tabular}{lc|R|bb|gg|GGG}
	\toprule
	\multicolumn{2}{c|}{ \TwoRowCell{ Performance \\ Indeces }}
										& N5		& N6		& \TwoRowCell{ MLP1 \\ Mean (Variance) }	& L			& \TwoRowCell{ MLP2 \\ Mean (Variance) }	& ODNP		& PVGM		&  CCD		\\
	\midrule
	\parbox[t]{2mm}{\multirow{9}{*}{\rotatebox[origin=c]{90}{DA Forecast}}} %
	&$RMSE$ (\si{\kilo\watt})			& $109$		& $110$		& \TwoRowCell{ $133$\\ $(2.05)$}			& $110$		& \TwoRowCell{$127$ \\ $(19.0)$}			& $227$ 	& $283$		& $121$ 	\\
	&$MAPE$								& $50.0\%$	& $50.7\%$	& \TwoRowCell{ $64.6\%$\\ $(5.17)$}			& $48.8\%$	& \TwoRowCell{$66.9\%$ \\ $(31.8)$}			& $87.4\%$ 	& $165.3\%$	& $61.4\%$	\\
	&$MBE$ (\si{\kilo\watt})			& $-1.275$	& $4.108$	& \TwoRowCell{ $-3.42$ \\ $(1.38)$} 		& $4.455$	& \TwoRowCell{$-1.15$ \\ $(37.7)$}			& $27.2$	& $127.4$ 	& $11.2$	\\
	&$R^2$								& $0.828$	& $0.827$	& \TwoRowCell{ $0.746$ \\ $(\num{2.98e-5})$}& $0.825$	& \TwoRowCell{$0.770$ \\ $(\num{2.71e-4})$} & $0.254$	& $-0.157$ 	& $0.787$	\\
	&$NRMSE$							& $0.414$   & $0.416$   & \TwoRowCell{ $0.511$ \\ $(\num{2.94e-5})$}& $0.418$	& \TwoRowCell{ $0.484$ \\ $(\num{2.28e-4})$}& $0.864$	& $1.076$	& $0.462$	\\
	&$RMSE_{NP}$						& $0.119$	& $0.119$	& \TwoRowCell{ $0.146$ \\ $(\num{2.41e-6})$}& $0.120$	& \TwoRowCell{ $0.139$ \\ $(\num{1.88e-5})$}& $0.248$	& $0.308$	& $0.132$	\\
	&$MAPE_{NP}$						& $7.51\%$	& $7.78\%$	& \TwoRowCell{ $9.85\%$ \\ $(\num{1.70e-2})$}& $7.71\%$	& \TwoRowCell{ $9.28\%$ \\ $(\num{1.24e-1})$}& $15.28\%$& $23.96\%$	& $9.49\%$	\\	
	\midrule
	\parbox[t]{2mm}{\multirow{9}{*}{\rotatebox[origin=c]{90}{HA Forecast}}} %
	&$RMSE$ (\si{\kilo\watt})			& $103$		& $104$		& \TwoRowCell{$129$ \\ $(6.79)$}			& $105$		& \TwoRowCell{$125$ \\ $(14.4)$}			& -			& $223$		& $120$		\\
	&$MAPE$								& $45.1\%$	& $46.7\%$	& \TwoRowCell{$66.9\%$ \\ $(11.5)$}			& $42.7\%$	& \TwoRowCell{$75.8\%$ \\ $(183)$}			& -			& $177.3\%$ & $61.4\%$	\\
	&$MBE$ (\si{\kilo\watt})			& $11.74$	& $11.97$	& \TwoRowCell{$-2.03$ \\ $(2.63)$}			& $12.53$	& \TwoRowCell{$-4.01$ \\ $(45.1)$}			& - 		& $62.9$	& $24.8$	\\
	&$R^2$								& $0.846$	& $0.841$	& \TwoRowCell{$0.756$ \\ $(\num{9.73e-5})$}	& $0.839$	& \TwoRowCell{$0.774$ \\ $(\num{1.89e-4})$}	& -			& $0.274$	& $0.788$	\\
	&$NRMSE$							& $0.393$	& $0.399$	& \TwoRowCell{$0.498$ \\ $(\num{9.88e-5})$} & $0.401$	& \TwoRowCell{$0.477$ \\ $(\num{1.77e-4})$} & -			& $0.852$	& $0.460$	\\
	&$RMSE_{NP}$						& $0.112$	& $0.114$	& \TwoRowCell{$0.142$ \\ $(\num{8.03e-6})$} & $0.115$	& \TwoRowCell{$0.136$ \\ $(\num{1.44e-5})$} & -			& $0.243$	& $0.131$	\\
	&$MAPE_{NP}$						& $6.88\%$	& $7.07\%$	& \TwoRowCell{$9.05\%$ \\ $(\num{3.28e-2})$}& $7.16\%$	& \TwoRowCell{$8.83\%$ \\ $(\num{5.28e-2})$}& -		 	& $18.66\%$	& $9.73\%$	\\
	\midrule
	%
	\end{tabular}
	\caption{Performance comparison of parametric models, ODNP, PVGM, CCD and MLP computed starting from day $57$. Results are grouped by model complexity and purpose in order to simplify the comparison: $5$-parameter model (N5) in red column, $6$-parameter models (N6 and MLP1) in blue columns, $11$-parameter models (L and MLP2) in green column, and benchmark models (ODNP, PVGM and CCD) in gray columns.}
	\label{TAB:MLP_CMP_ONLINE}
\end{table}

\begin{figure*}[htbp!]
	\psfrag{day}{\hspace*{-7mm}\tiny{time (day)}}
	\psfrag{kW}{\hspace*{-5mm}\tiny{$RMSE_d$ (\si{\kilo\watt})}}
	\centering
	\includegraphics[width=1\textwidth]{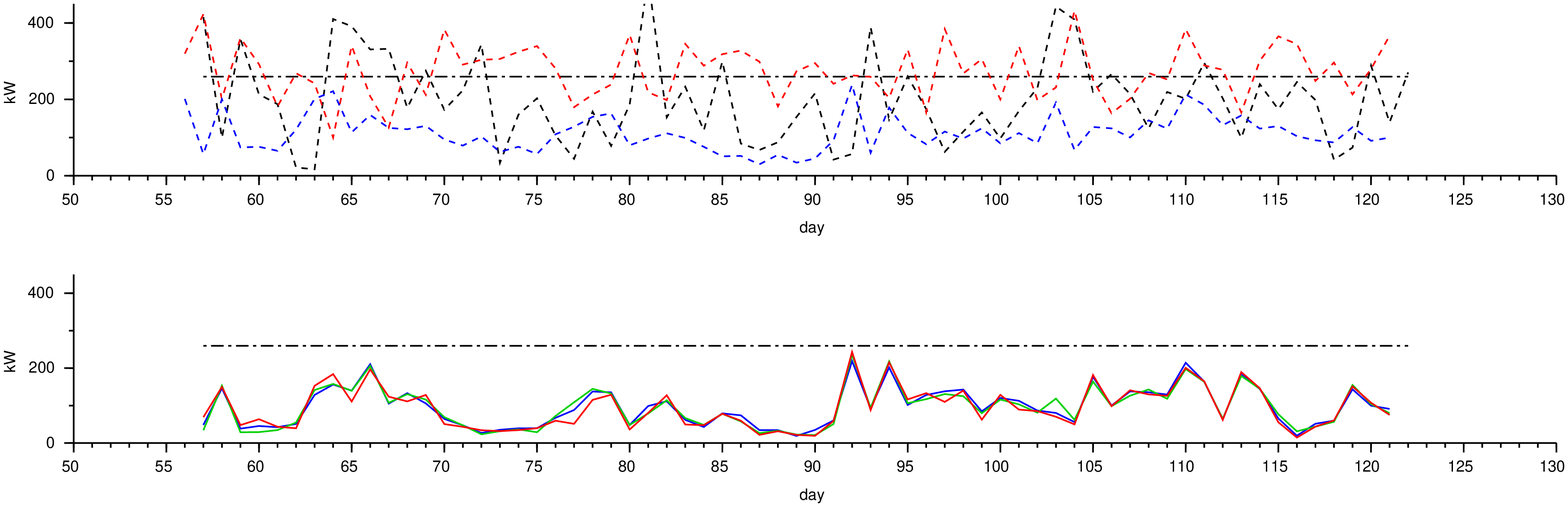} \\
	\caption{Comparison between standard deviation of the measured power (dashed dotted line) and the $RMSE_d$ computed on DA~forecasts. Top: ODNP (dashed line), PVGM (dasehd red line) and CCD (dashed blue line). Bottom: L model (green line), N5 model (red line) and N6 model (blue line).}	
	
	\label{FIG:RMSE}
\end{figure*}

\begin{figure*}[htbp!]
	\psfrag{day}{\hspace*{-7mm}\tiny{time (day)}}
	\psfrag{kW}{\hspace*{-5mm}\tiny{$RMSE_d$ (\si{\kilo\watt})}}
	\centering
	\includegraphics[width=1\textwidth]{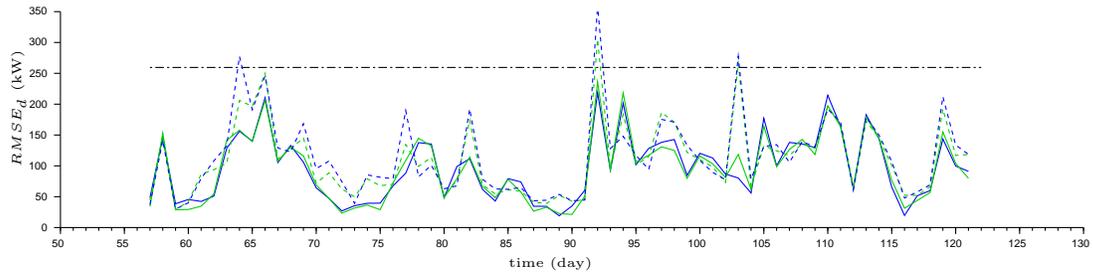} \\
	\caption{DA forecast. Comparison between standard deviation of the measured power (dashed dotted line), the $RMSE_d$ computed using the N6 model (blue line), MLP1 (blue dashed line), L model (green line) and MLP2 (green dashed line).}
	\label{FIG:RMSE_CMP1}
\end{figure*}

\begin{figure*}[htbp!]
	\psfrag{day}{\hspace*{-7mm}\tiny{time (day)}}
	\psfrag{kW}{\hspace*{-5mm}\tiny{$RMSE_d$ (\si{\kilo\watt})}}
	\centering
	\includegraphics[width=1\textwidth]{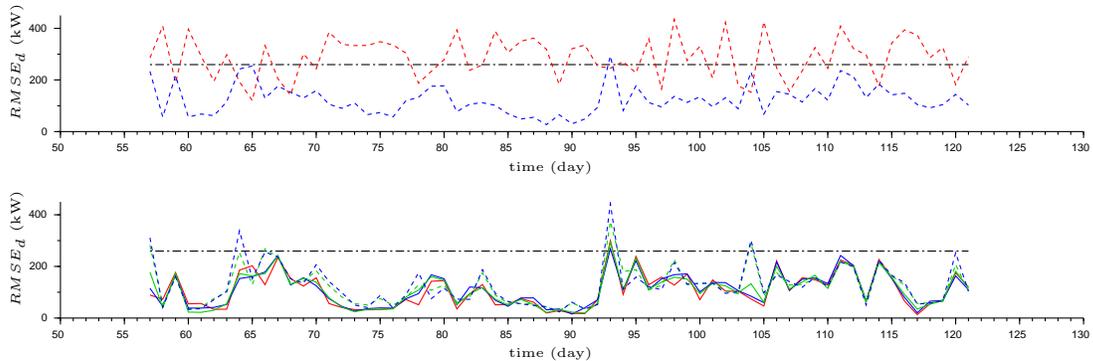} \\
	\caption{HA forecast. Comparison between the standard deviation of the measured power (dashed dotted line) and the $RMSE_d$. Top: PVGM (dasehd red line) and CCD (dashed blue line). Bottom: computed using the N5 model (red line), N6 model (blue line), MLP1 (blue dashed line), L model (green line) and MLP2 (green dashed line).}
	\label{FIG:RMSE_CMP2}
\end{figure*}

Figure~\ref{FIG:RMSE} shows the daily RMSE ($RMSE_d$) achieved by the proposed models in the DA forecast, compared with ODNP, PVGM, and CCD. In Figures~\ref{FIG:RMSE_CMP1}, such error is compared with those achieved by the MLPs in the DA forecast. Figure \ref{FIG:RMSE_CMP2} depicts the comparison of the $RMSE_d$ achieved by the various models and benchmarks in the HA forecast. In the above figures, all errors are also compared with the standard deviation of measured power data. 

Both in Table \ref{TAB:MLP_CMP_ONLINE} and Figure \ref{FIG:RMSE_CMP2}, data for the ODNP in the HA case are not reported as the ODNP is unsuitable as a reference model.

The apparently high normalized forecast errors (in particular the MAPE) are mainly due to the fact that the CCI data refer to a weather station located about \SI{20}{\kilo\metre} away from the plant site. Another source of forecast dispersion is the rough CCI data resolution. Both these two sources of uncertainty deteriorate forecasting accuracy as much as the cloudiness at the plant site changes rapidly in time. It is worth remarking, however, that the proposed models use a very limited amount of information which is easily and cheaply available, and that such raw data is informative enough for a reliable estimation of the generation model, as demonstrated by the parameter convergence properties highlighted in Figure \ref{FIG:RP_PARAMETER6}. Furthermore, it is expected that the proposed approach can be fruitfully adapted to large-scale aggregation of plants over macro areas, yielding significant performance improvements with respect to the single plant case \cite{WIDEN2015}, with no additional data requirements (except generation measurements). 


\begin{figure}[htbp!]
	\power
	\includegraphics[width=\textwidth]{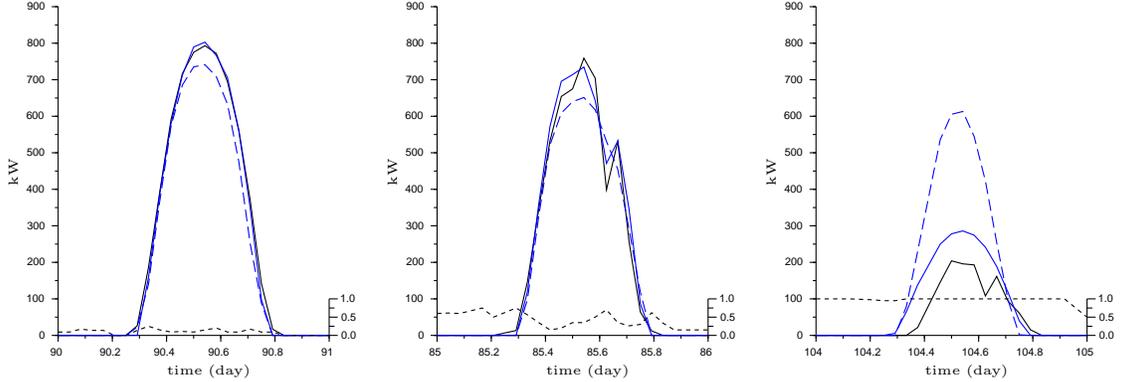}
	\caption{DA forecast. Comparison between the measured power (black line), N6 model forecast (blue line), and MLP1 forecast (dashed blue line). Dashed dotted line is the measured CCI. From right to left,  an almost clear-sky day, an overcast day and a uniformly overcast day are depicted.}
	\label{FIG:FORECAST_NOANGLE}
\end{figure}

\enlargethispage{-\baselineskip}
In Figure~\ref{FIG:FORECAST_NOANGLE} the DA forecasts provided by N6 model and MLP1 during three different days and under three different weather conditions is compared with the measures of generated power.

\enlargethispage{-1\baselineskip}
The procedure has also been tested using the last two weeks of data for validation purposes only, that is, model parameter values were frozen two weeks before the end of the data set, then offline DA and HA forecasts for the whole two following weeks were generated according to the modalities above, and the respective performance indices were computed. It turned out that the performances of the various models in terms of all indices were almost identical to those obtained with online parameter adaptation (the difference in terms of all error measures was below 1\%). In this respect, it is worth remarking that online parameter estimation has the advantage of capturing seasonal parameter variations (especially as far as $C(N)$ is concerned).

As far as the computational time is concerned, the MLP update algorithms take generally longer to execute with respect to the parameter update procedure. More specifically, a MLP training step using a $K$-sized sample data set requires longer computational time than $K$ model parameter updates using a single data sample. The overall computational times for all models and MLPs are reported in Table~\ref{TAB:MLP_COMP}. All algorithms were implemented in Scilab \cite{BIB:SCILAB} version 5.5.2 and executed on a 2.4 Ghz Intel Xeon(R) v3 processor running the Linux operating system. 
\begin{table}[htbp!]
	\centering
	\begin{tabular}{c|R|bb|gg|GG}
	\toprule
	{\scriptsize \begin{tabular}{@{}c@{}}Performance \\ Indices\end{tabular}}
										& N5	& N6	& \TwoRowCell{MLP1\\Mean (Variance)}& L			& \TwoRowCell{MLP2\\Mean (Variance)}	& PVGM 	& CCD		\\
	\midrule
	Sim. Time (\si{\second})			& $7.87$& $7.86$& \TwoRowCell{$127$ \\ $(285)$}		& $7.58$	& \TwoRowCell{$130$\\$(433)$} 			& $9.58$& $10.22$	\\			
			
	\bottomrule
	\end{tabular}
	\caption{Computational times. Results are grouped by complexity and purpose in order to simplify the comparison: $5$ parameters (N5) in red column, $6$ parameters (N6 and MLP1) in blue columns, $11$ parameters (L and MLP2) in green column, and benchmark models (ODNP, PVGM and CCD) in gray columns.}\label{TAB:MLP_COMP}
\end{table}
Summarizing the discussion above, we found that the proposed parametric models show comparable forecasting behavior and a general performance improvement with respect to MLPs, both from an accuracy and a computational viewpoint. In particular, model N6 exhibits the best compromise between performance, complexity and convergence/consistency properties. This apparently shows that a slight overparameterization with respect to a minimal model inspired by the theoretical models of CCF and power generation, turns out to be beneficial. On the contrary, excessive overparameterization impacts model identifiability and consistency.

\section{Conclusions}\label{SEC:CONC}
In this paper, we have proposed an efficient parametric technique aimed at model estimation for direct forecasting of PV power generation using cloud cover data. The approach is based on recursive least squares and Extended Kalman Filter. To this purpose, three models have been singled out: an 11-parameter linear model, and two nonlinear models, involving 5 and 6 parameters, respectively. The procedure is especially fit for the typical scenario where the network operator, due to the large number of managed producers, has no access to on-site irradiance and temperature measurements. The method exploits only the historical time series of generated power, cloud cover, and forecast temperature, which can be obtained from a meteorological service. The procedure involves modest memory and computational requirements. Its performance has been evaluated on simulated data as well as on a single plant located in Italy. The experimental results show a substantial improvement in terms of typical error measures with respect to alternative approaches based on Neural Networks and standard linear autoregressive models. All the proposed models show equivalent forecasting performance, while the 6-parameter nonlinear model generally outperforms the other two as far as data fitting properties are concerned. It is expected that the proposed approach can be fruitfully adapted to the aggregation of several plants over macro areas, with significant performance improvements with respect to the single plant case. This issue is the subject of current investigation.

\section*{References}
\bibliographystyle{elsarticle-num}
\bibliography{../biblio}

\end{document}